\documentclass[runningheads]{llncs}
\usepackage[utf8]{inputenc}
\usepackage{graphicx}
\usepackage{tabularx}
\usepackage{comment}
\usepackage{booktabs}
\usepackage{caption}
\usepackage[colorinlistoftodos]{todonotes}
\usepackage{paralist}

\begin{document}
\title{Event Log Generation: An Industry Perspective} 
%
%
\author{Timotheus Kampik\inst{1}\orcidID{0000-0002-6458-2252} \and \\
Mathias Weske\inst{2}\orcidID{0000-0002-3346-2442}\thanks{Work by this author supported by the SAP Academic Fellowship program.}
}
\authorrunning{Kampik and Weske}
%
\institute{SAP Signavio, Berlin, Germany \\
\email{timotheus.kampik@sap.com}\\
\and
Hasso Plattner Institute, University of Potsdam, Germany \\
\email{mathias.weske@hpi.de}}
\maketitle              
\begin{abstract}
This paper presents the results of an industry expert survey about event log generation in process mining. It takes academic assumptions as a starting point and elicits practitioner's assessments of statements about process execution, process scoping, process discovery, and process analysis. The results of the survey shed some light on challenges and perspectives around event log generation, as well as on the relationship between process models and process execution, and derive challenges for event log generation from it. The responses indicate that particularly relevant challenges exist around data integration and quality, and that process mining can benefit from a systematic integration with more traditional and wide-spread business intelligence approaches.
\end{abstract}
\section{Introduction}
\label{intro}
Started as an academic discipline, the focus of process mining has mostly been on concepts and algorithms that analyze observed process behavior and compare it to behavior that has been defined in process models. Process mining is based on event logs, which represent real-world business process executions. More concretely, an event log consists of a sequence of events, each of which includes at least a case identifier and an activity reference. Until recently, the assumption has been that event logs of this structure are readily available~\cite{DBLP:journals/widm/DibaBWW20}. With the industrial uptake of process mining, this assumption has been challenged, and the importance of event log generation has become evident. Practical experiences indicate that the generation of event logs incurs substantial efforts~\cite{eventlogsDatabases,ANDREWS2020113265}.

In order to better understand the practical challenges of event log generation in process mining, the authors have conducted a survey with different stakeholders ranging from domain experts to systems designers and software engineers. This paper describes the structure of the survey as well as the main results of the empirical study, and it derives focal areas for industrially relevant research in event log generation, and -- more broadly -- in process mining.

The remainder of this paper is organized as follows. After introducing the main concepts in process mining, we motivate the survey and provide the research questions that we aim to answer. The survey is presented and its results are discussed, before concluding remarks complete the paper.

\section{Process Mining Overview}
\label{pm:overview}
\begin{figure}[t]
    \centering
    \includegraphics[width=0.85\linewidth]{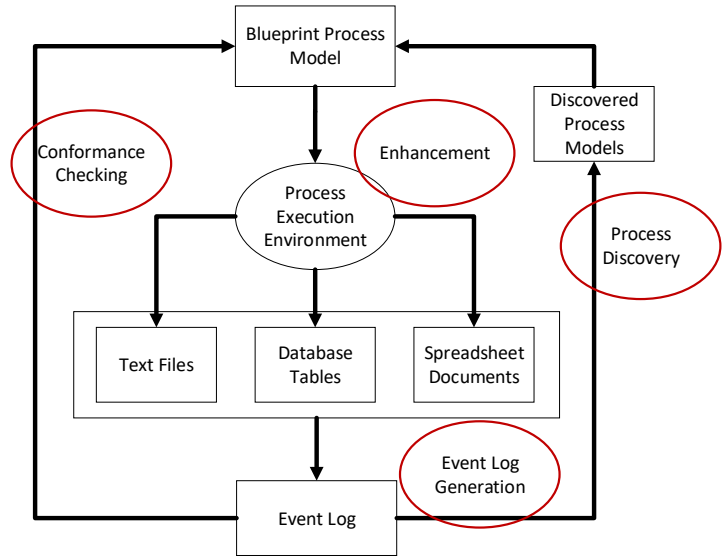}
    \caption{Overview of process mining concepts and tasks}
	\label{fig:overview}
\end{figure}

For more than a decade, the academic process mining community has developed an impressive arsenal of process mining methods, techniques, and tools. Several of those have recently found their way to industrial practice. In this section, the main tasks that can be performed in process mining projects are categorized and the role of event log generation is highlighted. In Figure~\ref{fig:overview} important concepts in process mining are presented, and the different process mining tasks are shown.

Business process management is based on process models that provide an abstract representation of the business processes of an organization. These process models are used in different ways. So-called \emph{as-is} process models describe the current state of the business processes. They are analyzed and improved, leading to \emph{to-be} process models, which represent new and improved business processes that will be implemented in the organization. Since they acts as blueprints for business processes, we refer to them as ``blueprint process model(s)'' in Figure~\ref{fig:overview}.

A blueprint process model is used to implement the corresponding business process, configuring a process execution environment as specified in the process model. It is worth noting that processes are not always implemented exactly as specified by the blueprint process model. In fact, finding and quantifying deviations of the blueprint process model and the executed process instances is one important challenge in process mining. Blueprint process models can even be missing, so that the process is implemented based on a traditional requirements engineering effort, or merely based on the understanding that the systems engineers have of the process.

During process execution, process execution data is generated. This data is located in different data stores. It can be of an arbitrary structure, ranging from well-structured data stored in relational databases to spreadsheets and text files. Event log generation is facing the challenge of integrating those data sources and to provide an appropriate basis for the main process mining tasks \emph{process discovery} (and other analyses), \emph{conformance checking}, and \emph{process enhancement}.

In process discovery, the event log is used to ``re-engineer'' a process model whose instances have been observed in the event log. Ideally, the generated pro- cess model matches the blueprint process model. As experience shows, this is hardly ever the case. Conformance checking provides techniques and tools to compare an event log with the blueprint process model. Valuable insights can be inferred from conformance checking, for instance about missing activities or activities that have been observed in the event log and that cannot be found in the blueprint process model. Obviously, conformance checking can only be performed if a blueprint process model is available. In process enhancement, we analyze the discovered process model and draw conclusions on how to improve -- or, enhance -- the blueprint process model. This task is typically performed by domain experts that are knowledgeable about the process and can interpret the discovered process model properly.

\section{Motivation, Research Questions, and Survey Structure}
\label{methods}
While at first sight event log generation seems like a straightforward task, in practice, it turns out that it is not. The complicating factors are manifold, ranging from heterogeneous data sources and data quality issues to challenges that are related to the goal of the process mining task at hand. 

To motivate challenges in process mining and their effect on event log generation, we consider a process mining project in a hospital setting that looks at medication aspects of patients with lower back pain syndrome. An important aspect of event log generation is the definition of a case identifier. In the hospital example, we might choose the patient identifier as case identifier. This is convenient, since the data entries that we find always have a patient identifier associated. However, selecting the patient identifier as case identifier might lead to undesired outcomes during process analysis. This is due to the fact that a case does not only contain activities that are related to lower back pain, the disease we are interested in. Instead, other diseases that the patient has suffered from are also part of the process that we mine (and its instances). If a patient suffered from an arm fracture, she might have been administered painkillers already, which would be falsely associated with her lower back pain condition that we are interested in. This example illustrates that the selection of the case identifier has severe implications on event log generation.

Based on these considerations, the survey aims at answering the following research questions.
\begin{description}
     \item[RQ1:] How do process models impact process executions and how are process mining opportunities affected?
     \item[RQ2:] What are the main conceptual challenges in event log generation?
     \item[RQ3:] What are the main technical challenges in event log generation?
\end{description}
To provide a holistic view on the different challenges of event log generation and to address the research questions, the following areas are covered by the survey.
\begin{itemize}
    \item Process Execution: Since event logs are based on data that is generated during business process execution, it is essential to investigate how business processes are actually executed. This area of questions involves the role of process models as well, because those are required for specific process mining tasks, for instance for conformance checking~\cite{DBLP:books/sp/Aalst16}.
    \item Process Scoping: If events in an event log belong to different business processes and we use that log in process discovery, the resulting process model becomes complex and does not reflect the desired process properly. One reason is improper scoping of the process, which is another important aspect that needs to be covered when generating event logs~\cite{DBLP:books/sp/15/Aalst15}. 
    \item Data Sources and Event Logs: Event logs use data that might be stored in different, heterogeneous data sources~\cite{DBLP:journals/widm/DibaBWW20}. The questions in this area address the quality and number of data sources used in process mining projects as well as the effort that is incurred by event log generation.
    \item Process Discovery: In process discovery, we are interested in discovering process models from event logs. We can compare those discovered process models with process models that have served as blueprints of the process execution. To find out more about these aspects, the survey contains questions related to issues in process discovery that might point to problems related to the event log that was used as input to process discovery.
    \item Process Analysis: Even though process discovery can be regarded as a subset of process analysis, we have decided to separate these two areas. In process analysis, we ask questions related to performance indicators of the process. The questions are important for event log generation, since we have to make sure that the relevant data attributes actually find their way to the events in the generated event log.
\end{itemize}
In each area mentioned, the survey asked for the assessment of several statements, such as ``It is straightforward to find the correct scope of a process, from start to end.''. The survey invited answers based on a 5-point Likert scale (\emph{Strongly disagree/SD}, \emph{Disagree/D}, \emph{Neutral/N}, \emph{Agree/A}, and \emph{Strongly Agree/SA}). The area \emph{Data Sources and Event Logs} was augmented by two additional questions about the typical number of data sources of an event log and common types of data sources. Broader open text feedback could be provided as well, but is subject to organizational nondisclosure requirements\footnote{Let us highlight that no open feedback that contradicts the other survey results was received.}.

\section{Survey Results}
\label{results}
The authors have conducted the survey from December 2021 through January 2022. Employees of a large enterprise systems vendor have been asked to participate, and respondents were sampled from teams of process mining and business process intelligence experts. Subjects have different educational and professional backgrounds, from technical and engineering to business and management. They also serve in different roles in the company, including solutions, engineering, and product innovation. In the remainder of this section, we focus on the demographics of subjects, before considering the responses to the survey questions.

Prior to the survey, a pilot survey was conducted to gather feedback from five selected experts; the refinement based on the feedback resulted in the presented survey.
For each Likert-scale assessment, a bar chart with the responses is provided, alongside a table that provides (for the overall group of respondents, as well as for demographic groups) the median, mode, and a simplified mode (\emph{Sim. Mode}) that aggregates ``strongly disagree'' (SD) and ``disagree'' (D) to ``disagree'' (D), as well as ``strongly agree'' (SA) and ``agree'' (A) to ``agree'' (A).
One question asked for an approximate quantification (as a categorical answer/selection).
The content of one free-text answer is aggregated and summarized.

Because the differences between demographic groups are not the main focus of the study, and because it was not possible to control for confounding features like team-level organizational assignment or role changes over time, no analysis of the statistical significance of the assessment differences between demographic groups is made. We merely observe that general alignment with respect to the assessment direction typically brings with it alignment between demographic groups. In contrast, investigating the demographic impact on the lack of alignment as observed in the assessments of some statements is out of the scope of this paper and would require further research.

\subsection{Demographics}
Overall, the survey was answered by 32 subjects. Demographic information can be summarized as follows.
\begin{description}
     \item[Years of industry work experience:] 2 subjects (6.3\%) reported 0-1 years; 4 (12.5\%) 1-3 years; 11 (34.4\%) 3-5 years; 2 (6.3\%) 5 - 10 years; 13 (40.6\%) more than 10 years.
     \item[Years of process mining work experience:] 3 subjects (9.4\%) reported 0-1 years; 15 (46.9\%) 1-3 years; 13 (40.6\%) 3-5 years; 1 (3.1\%) 5-10 years.
     \item[Educational background:] The survey offered a range of options, as well as an open text field to specify the educational background. Aggregated, the categories \emph{Science/Engineering}: 23 (71.9\%) and \emph{Mixed/other}: 9 (28.1\%) were obtained.
     \item[Role in the organization:] the survey provided a selection of prevalent internal roles, as well as an open text field. The results were then aggregated into the categories \emph{Product / Engineering} (abbreviated as Pro./eng.): 15 (46.9\%) and \emph{Solutions / Consulting} (Sol./cons.): 17 (53.1\%).
\end{description}
One respondent reported more process mining work experience than industry work experience, which can potentially be explained by work experience in a non-industry context, such as in academia. As another aggregated category, the experience levels are aggregated to \emph{Experienced} (Exp) (14, 43.8\%) and \emph{Newcomers} (New) (18, 56.2\%), where falling into the former category requires at least three years of process mining experience, as well as at least five years of industry experience.

\subsection{Process Execution}
The to-be-assessed statements regarding process execution aimed at eliciting a broader, nuanced perspective on the roles that (formal) process models play in business process execution. Assessments of the following statements were requested.
\begin{enumerate}
    \item \emph{Business processes are executed exactly as specified in process models} (Table~\ref{fig:chart_11} in the Appendix).
    This statement reflects the traditional academic assumption that process models are executable specifications. Not surprisingly, most respondents disagreed (14) or strongly disagreed (12) with this statement, while merely two respondents agreed (one of the two strongly agreed)\footnote{Here and henceforth, the number of \emph{neutral} responses can -- if not explicitly stated -- be determined by subtracting the number of all other respondents from $32$.}. No substantial differences between demographic groups seem to exist.
    \item \emph{Process models are used as requirements specifications that are then implemented in IT systems} (Table~\ref{fig:chart_12}). This statement can be considered a relaxation of the previous one: if process models are not ‘directly’ executed, they at least inform the specification of systems that execute business processes. There is no consensus about this statement, but a simple majority (14: 10 $A$, 4 $SA$) of the respondents agreed with the statement, while relatively few (7) disagreed (no one strongly disagreed). Consultants reported to agree more with this statement than product managers and engineers; the same applies to respondents with technical education vs. ‘other/mixed’ education and experienced practitioners vs. newcomers.
    \item \emph{Process models are not used to implement processes} (Table~\ref{fig:chart_13}). This statement can be considered a contradiction of the former statement. Indeed, no respondent expressed agreement with this statement \emph{and} with the former statement. Most respondents disagreed (6) or strongly disagreed (15), whereas only two respondents agreed (no one strongly agreed). The assessment is consistent across demographic groups.
\end{enumerate}
The responses suggest that process models are rarely directly executed, but still relevant for execution in that they inform business process implementation in IT systems in some ways. While this conclusion is not particularly surprising from an industry perspective, it allows for the conclusion that academically it is important to acknowledge that many process models are primarily for humans to understand and discuss and not necessarily for machines to automatically execute.

%
%

%
%

\subsection{Process Scoping and Data Sources}
The statements regarding process scoping aimed at eliciting an assessment of how challenging the identification of events and the data sources that provide them actually is. Assessments of the following statements were requested.
\begin{enumerate}
    \item \emph{It is straightforward to find the correct scope of a process, from start to end} (Table~\ref{fig:chart_21}). This statement challenges the assumption that identifying the scope of a process, from start to end, is indeed challenging. Most respondents strongly disagreed (7) or disagreed (15) with this statement; a small minority of respondents (3) agreed with the statement (no one strongly agreed). Disagreement is consistent across demographic groups.
    \item \emph{It is straightforward to group events to process instances (finding the case ID, group by case ID)} (Table~\ref{fig:chart_22}). This statement challenges the assumption that event correlation is challenging. Respondents broadly disagreed (15) or strongly disagreed (5) with the statement; however there is some misalignment among respondents, with seven respondents reporting agreement and two strong agreement. Still, disagreement is dominant across demographic groups.
    \item \emph{It is straightforward to locate the data sources that we need for generating an event log} (Table~\ref{fig:chart_23}).
    The statement claims that locating data sources for event log generation is trivial. While most respondents strongly disagreed (3) or disagreed (13) with this claim, there are also some who agreed (7) or strongly agreed (1). Overall, the median is between disagreement and a neutral attitude. We find differences among the demographic groups. The median is ‘disagree’ for respondents who work in product development, as well as for respondents who have a not exclusively technical education and respondents who are relatively new to process mining or industry work. It is ‘neutral’ for respondents with technical education, respondents who work in solutions/consulting, and respondents who are generally more experienced. While some form of disagreement is the most common response type across all demographic groups, the responses are largely inconclusive: locating data sources may be a challenge, but is not necessarily so.
    \item \emph{Typically, data quality issues do not affect event log generation} (Table~\ref{fig:chart_24}).
    The statement challenges the practical assumption that a key problem in process mining is obtaining high-quality data and mitigating data quality issues. Most respondents strongly disagreed (13) or disagreed (12) with this statement, while there was little agreement (3) and strong agreement (1). Disagreement dominates across demographic groups, suggesting that addressing issues around data quality is in fact a challenge when generating event logs.
\end{enumerate}
The responses suggest that scoping business processes regarding their temporal scope (from start to end), as well as correlating events to cases (identifying case IDs) is challenging. Also, data quality issues are prone to affect event log generation. It could not be confirmed that the identification of data sources poses a substantial challenge.

\subsection{Event Logs}
\begin{figure}[!ht]
    \centering
    \includegraphics[width=0.85\linewidth]{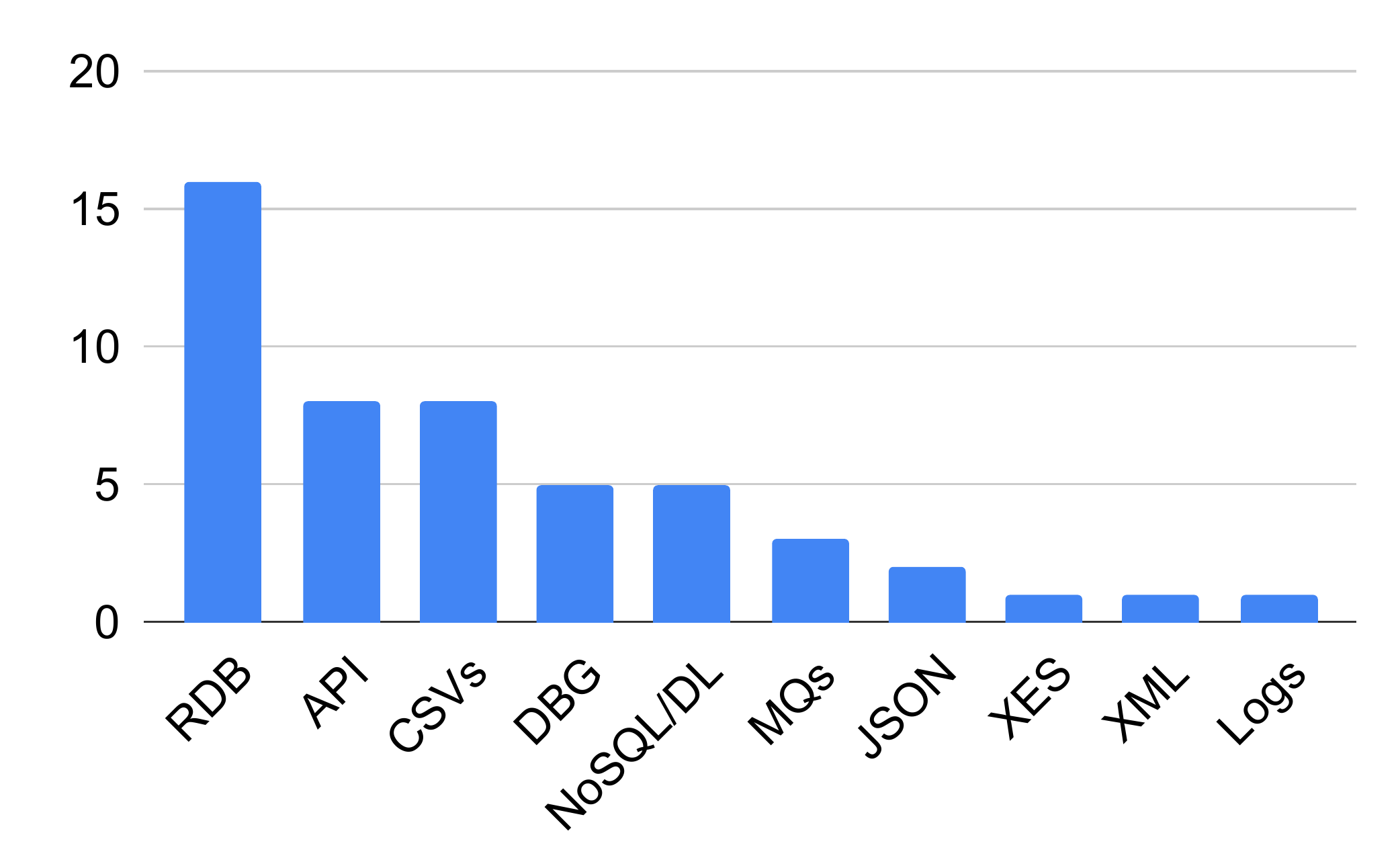}
    \caption{Data sources for event log generation.}
	\label{fig:data-soruces}
\end{figure}

The requests for assessment were augmented with additional questions about data sources for event logs. Assessments of the following statements/answers to the following questions were requested.
\begin{enumerate}
    \item \emph{Event log generation incurs significant efforts in process mining projects} (Table~\ref{fig:chart_31}).
    This statement reflects the assumption that a substantial part of overall efforts in process mining are spent on event log generation. Most respondents strongly agreed (13) or agreed (13) with this statement. Merely 2 respondents disagreed (and no one strongly disagreed). Agreement is largely consistent across demographic groups.
    \item \emph{Extract-Transform-Load (ETL) pipelines provide all information needed in an event log} (Table~\ref{fig:chart_32}).
    The statement asks for an assessment of the extract-transform-load pipeline architecture for event log generation. A simple majority of respondents assessed the statement as neutral (13), whereas eight respondents agreed, ten disagreed and one strongly disagreed (no one strongly agreed). Disagreement is somewhat stronger among respondents with a product/engineering background, as well as among respondents with a non-technical or hybrid education. Overall, no clear signal of support or opposition to the statement could be elicited.
    \item \emph{How many backend systems are typically providing the data for a single event log?} (Table~\ref{fig:chart_33}).
    The statement challenges the assumption that event logs are typically generated from the data provided by a single system. Generally, there is no agreement on how many systems are typically used; six respondents stated that one system is used (6/``one''); otherwise the responses are: (8/``two''), (1/``three''), (9/``more than three''), and (8,``I don’t know''). The relatively large proportion of participants that answered ``I don’t know'' can potentially be explained by the fact that some respondents wanted to indicate that there is no simple answer, i.e., the number of systems varies between projects. Interestingly, respondents who work as consultants most frequently stated that typically, one backend system is used, whereas respondents with an exclusively technical education equally frequently reported the use of one and of two backend systems; respondents with substantial industry and process mining experience most frequently selected ``two'' (however, equally many selected ``I don't know''); these modes are lower than the overall mode, and the modes provided by other demographic groups. The responses allow drawing the conclusion that event logs are not necessarily generated from a single backend system.
    \item \emph{In a given system, the information needed for an event log is stored in a single relational table} (Table~\ref{fig:chart_34}).   This statement somewhat naively asserts that event log data can be extracted from exactly one relational database table but does not directly contradict the previous statement (as it is possible that the data can be provided by several systems, but by exactly one table in each system). Most respondents strongly disagreed (17) or disagreed (10) with this statement; merely one respondent agreed with the statement (no one strongly agreed). Disagreement is largely consistent across demographic groups. The results suggest that typically, event logs cannot straightforwardly be extracted by reading out data from one specific database table.
    \item \emph{What are typical data sources of event logs? (E.g., relational database tables, document collections, CSVs, ...).} The question aims at getting an overview of typical data sources. After manually clustering the responses (considering that one participant can provide multiple responses as part of the free-text answer), the following categories are obtained and populated.
    \begin{inparaenum}[i)]
        \item relational databases and tables thereof (RDB): 16 respondents; 
        \item API access or similar to enterprise systems (API): 8;
        \item CSVs files (CSVs): 8;
        \item Database (generic, DBG): 5;
        \item no SQL/big data storages/data lakes (NoSQL/DL): 5;
        \item Message queues/event-based (MQS): 3;
        \item JSON content (JSON): 2;
        \item XES or XML files, or logs: 1 each.
    \end{inparaenum}
    Figure~\ref{fig:data-soruces} displays the categories and the number of responses that reflects each category.
\end{enumerate}

The responses confirm the assumption that event log generation efforts are substantial and that sources for event logs are often relational databases of enterprise systems or CSVs, whereas data lakes and event-based systems seem to be emerging as alternatives.
In contrast, XES is -- apparently -- typically not used (and possibly not available), which raises questions about the practical importance of the XES XML standard\footnote{This finding is to some extent confirmed by the results of another recent expert survey~\cite{10.1007/978-3-030-98581-3_1}.}.
Additionally, the responses suggest that event logs are typically not extracted from a single database table and not necessarily from a single backend system.

\subsection{Process Discovery}
The statements regarding process discovery aimed at gauging the relevance of mining complex control flows from an industry perspective. Assessments of the following statements were requested.
\begin{enumerate}
    \item \emph{In process discovery, we encounter complex process models} (Table~\ref{fig:chart_41}).  The statement asserts that process complexity plays a role in process discovery.
    Almost all respondents strongly agreed (18) or agreed (13) with this statement (no one disagreed or strongly disagreed). Agreement is largely consistent across demographic groups, with respondents working in product or engineering roles expressing slightly less strong agreement. This lets us conclude that managing complexity plays a relevant role in process discovery.
    \item \emph{In process discovery, the ordering of activities is important to me (e.g., it is important to know that “activity A always precedes activity B”)} (Table~\ref{fig:chart_42}).
    This statement asserts that the notion of a process as an ordered sequence of activities is practically relevant when discovering processes. Almost all respondents strongly agreed (14) or agreed (15) with this statement (no one strongly disagreed or disagreed). Across demographic groups, agreement strength varies, but there is generally clear agreement, which supports the conclusion that activity ordering is indeed important.
    \item \emph{Most discovered processes are sequential (no branching or concurrency)} (Figure/Table~\ref{fig:chart_43}).
    This statement challenges the assumption that process complexity in terms of variance and concurrency matters and stands in contrast to the first statement in this group. Most respondents strongly disagreed (8) or disagreed (14), whereas relatively few agreed (3) or strongly agreed (2). Disagreement is somewhat consistent across demographic groups, although respondents working in consulting exhibit a more neutral attitude. Disagreement is relatively strong among respondents with substantial experience or with a not exclusively technical background. The results allow us to carefully draw the conclusion that discovered processes tend to be not sequential. Interestingly, the signal we get from these responses is weaker than the signal that we get from the responses to the first statement, indicating that branching and concurrency may not account for all complexity that we encounter in process discovery.
\end{enumerate}
The responses suggest that managing complexity, activity ordering and (to a slightly lesser extent) variants, are indeed practically relevant challenges.

\subsection{Process Analysis}
The statements regarding process analysis aimed at eliciting an assessment of the relevance of Key Performance Indicators (KPIs), conformance checking, and traditional Business Intelligence (BI) in the context of process analysis. Assessments of the following statements were requested.
\begin{enumerate}
    \item \emph{An important goal of process mining is the calculation of KPIs} (Table~\ref{fig:chart_51}). The statement asserts that Key Performance Indicators (KIPs, or, in the context of business process management: Process Performance Indicators, PPIs) play an important role in process mining. Most respondents strongly agreed (10) or agreed (17) with this statement. Merely two respondents disagreed (no one strongly disagreed). Agreement is consistent across demographic groups, which lets us conclude that KPI calculation is indeed important.
    \item \emph{In process mining, it is difficult to identify meaningful KPIs} (Table~\ref{fig:chart_52}). The statement asserts that identifying meaningful KPIs is a challenge, which is a widely accepted premise for performance measures in general. There is broad disagreement among the respondents with respect to this statement. While no respondent strongly agreed, many (12) agreed, and many strongly disagreed (3) or disagreed (11). Respondents that work as consultants or have an exclusively technical background expressed more agreement than other demographic groups. The results merely allow for the conclusion that it is not clear whether identifying meaningful KPIs is difficult; it may be difficult in some scenarios and straightforward in others.
    \item \emph{Comparing the event log with a process model is important in process analysis} (Table~\ref{fig:chart_53}). This statement reflects the notion of conformance checking, which is a key aspect of academic research on process mining. Most respondents strongly agreed (9) or agreed (13), while few strongly disagreed (1) or disagreed (4). Agreement is largely consistent across demographic groups, which lets us conclude that comparing event logs with blueprint process models is indeed important.
    \item \emph{A better integration of Business Intelligence (BI) and process mining would be valuable} (Table~\ref{fig:chart_54}). This statement reflects the practical intuition that process mining and related analyses is related to business intelligence and hence should be integrated with it. Most respondents strongly agreed (11) or agreed (17) with the statement, whereas merely one disagreed (no one strongly disagreed). Agreement is largely consistent across demographic groups. The results conclude that the integration of BI and process mining is indeed a relevant frontier for research and innovation.
\end{enumerate}
The responses suggest that KPIs play an important role in process mining, and that the integration of business intelligence and process mining is a practically relevant research direction, but also that comparing event logs with manually created process models (which relates to the academic research field of conformance checking) appears to be an important aspect. Whether it is difficult or not to identify KPIs cannot be answered by our study.

\section{Discussion}
It is important to highlight that the survey's findings need to be seen in the light of its limitations. In particular, the survey was conducted among employees of a single enterprise software system vendor with strong expertise in process management and process mining. Our 'insider' access allowed for a precise targeting of potential respondents. Considering that respondents are i) from different parts of an organizational unit that has been recently (prior to the survey) created as the result of an acquisition and ii) relatively diverse given their experience levels and roles in the organization, the strong alignment of results across demographics suggests that many of the findings can potentially (but not necessarily) be generalized by broader follow-up studies. Such studies are relevant future re- search, considering the specific population that the survey sampled from, as well as the relatively small sample size. The remainder of this section discusses the key findings of the survey.

\subsection{Questioning Academic Assumptions}

The academic business process management and process mining community has traditionally close contacts to industry, which is evident given the many university spin-offs (startups) in the area and many collaboration projects between academia and industry. Still, the focus of academia and research is, by nature, different from the main objectives of industry organizations. While industry focuses on practical challenges that provide value to customers, academia's main interest is well-scoped, intellectually challenging problems that look for elegant solutions. To come up with those solutions, academic assumptions have to be made.

With this survey we could confirm some of those assumptions, while rejecting others. Traditionally, academia has assumed that process models are interpreted by process engines that would enact the process exactly as specified. More recently, academia is critical about this assumption, even challenging the value of process models.

The survey provides interesting findings in this regard. It rejects the idea that business process models are exact specifications of processes that run in the real-world. At the same time, process models provide significant value by their role in defining requirements during systems development.

It is worth noting that the finding questions the direct link between model and execution, as depicted by Figure~\ref{fig:overview}. We have to read this link as information flow, being used in a translation from model to executable process. This translation requires human interpretation, typically with the help of dedicated systems for enterprise system configuration. More broadly speaking, because process models may be useful in ways that diverge from academic assumptions, approaches to assessing their correctness need to be re-thought as well\footnote{To allude to the famous quote that ``[e]ssentially all models are wrong, but some of them are useful'', as commonly attributed to George Box.}.

\subsection{Relevance of Academic Research}

The results indicate that well-established research directions that are concerned with the mining of process control flow and variants therein, as well as with the comparison of expected (modeled) and factual (mined) flow are important from an industry perspective. These findings are to be interpreted carefully, i.e., there are process mining practitioners who believe in the importance, but industry experts that focus on traditional business intelligence or machine learning-based analytics may assess the corresponding statements differently -- or are not aware of their potential. 
Process complexity is regarded as an important problem, which might hint at the challenges in process scoping, discussed above. If processes are not well scoped, this means that events of different processes are used in process discovery. Since these processes might run independently from each other, events occur concurrently, leading to complex process structures.

\subsection{Emerging Research Directions}
The findings suggest that questions of particular importance evolve around data quality, event correlation, and the integration of event log-based process mining with traditional business intelligence.  While  recent research starts to address some of these challenges, in particular around data on-boarding and integration~\cite{eventlogsDatabases,Dijkman2020,ANDREWS2020113265}, as well as event correlation and object-centered process mining~\cite{10.1007/978-3-030-30446-1_1}, the free text feedback gathered from the survey points to largely unexplored questions, \emph{e.g.}, the aforementioned integration of business intelligence approaches into process mining and the use of models as tools for event log generation and process scoping.

\section{Conclusion}
\label{conclusion}
The survey results presented in this paper shed some light onto challenges around event log generation. In particular, the results allow drawing the following conclusions: \begin{inparaenum}[i)]
\item process models are typically not directly executed, but rather serve as input for enterprise software system specification and configuration, which is obvious from an industry perspective, but is potentially a useful insight for academia;
\item identifying process start and end, as well as event correlation is a challenge;
\item data quality issues have an impact on event log generation;
\item classical academic questions in process discovery about process complexity, activity ordering and process variants are practically relevant;
\item event log generation incurs indeed substantial effort and event logs are usually generated based on several relational database tables, and frequently based on data from several backend systems;
\item data sources for event log generation are most commonly traditional relational databases of enterprise systems whose content is sometimes transferred into CSV format as a ‘low tech’ export/import procedure, but event-based systems and data lakes are emerging as sources as well;
\item the mining of control flow is practically relevant, and so is the generation of KPIs and the integration of process intelligence and business intelligence.
\end{inparaenum}
As a broader conclusion, the survey results suggest that the role of process models in process mining and event log generation, but also generally in architectural perspective on the process management life-cycle, needs to be re-assessed. In particular, the results indicate that while the connection between designed blueprint process models and executed process instances is rather indirect than direct. Models \begin{inparaenum}[i)]\item play a role in process implementation, but not as strong of a role as often assumed by academia;  \item can be used to better inform process scoping and event correlation; \item can ideally combine knowledge-based and data-driven process insights. \end{inparaenum}

\newpage
\bibliographystyle{splncs04}
\bibliography{refs}

\begin{table}
\subsubsection*{Appendix - Tables and Figures \newline}
	\begin{minipage}{0.5\linewidth}
		\centering
		\includegraphics[trim={1.25cm 0 1.75cm 0},clip,width=55mm]{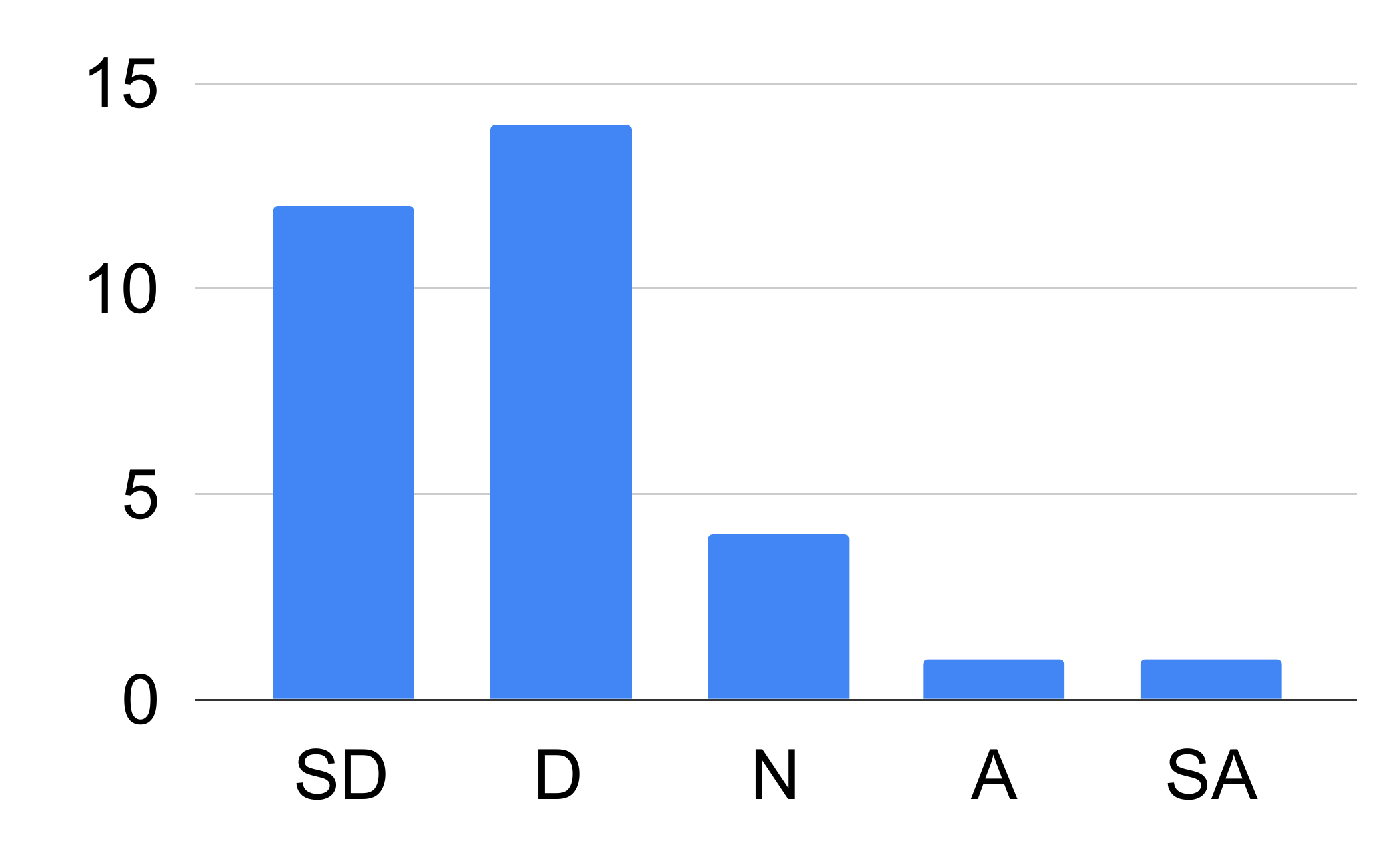}
	\end{minipage}
	\begin{minipage}{0.45\linewidth}
	    \footnotesize
		\centering
        \begin{tabular}{ | l | l | l | l | }
        \hline
        \textbf{Group} & \textbf{Median} & \textbf{Mode} & \textbf{Sim. Mode} \\ \hline
        All & D & D & D \\ \hline
        Pro./eng. & D & SD & D \\ \hline
        Sol./cons. & D & D & D \\ \hline
        Tech. ed. & D & D & D \\ \hline
        Other ed. & D & SD/D & D \\ \hline
        Exp. & D & D & D \\ \hline
        New & D & SD/D & D \\ \hline
      \end{tabular}
	\end{minipage}\hfill
	\caption{Results. \emph{Business processes are executed exactly as specified in process models}.}
	\label{fig:chart_11}

	\begin{minipage}{0.5\linewidth}
		\centering
		\includegraphics[trim={1.25cm 0 1.75cm 0},clip,width=55mm]{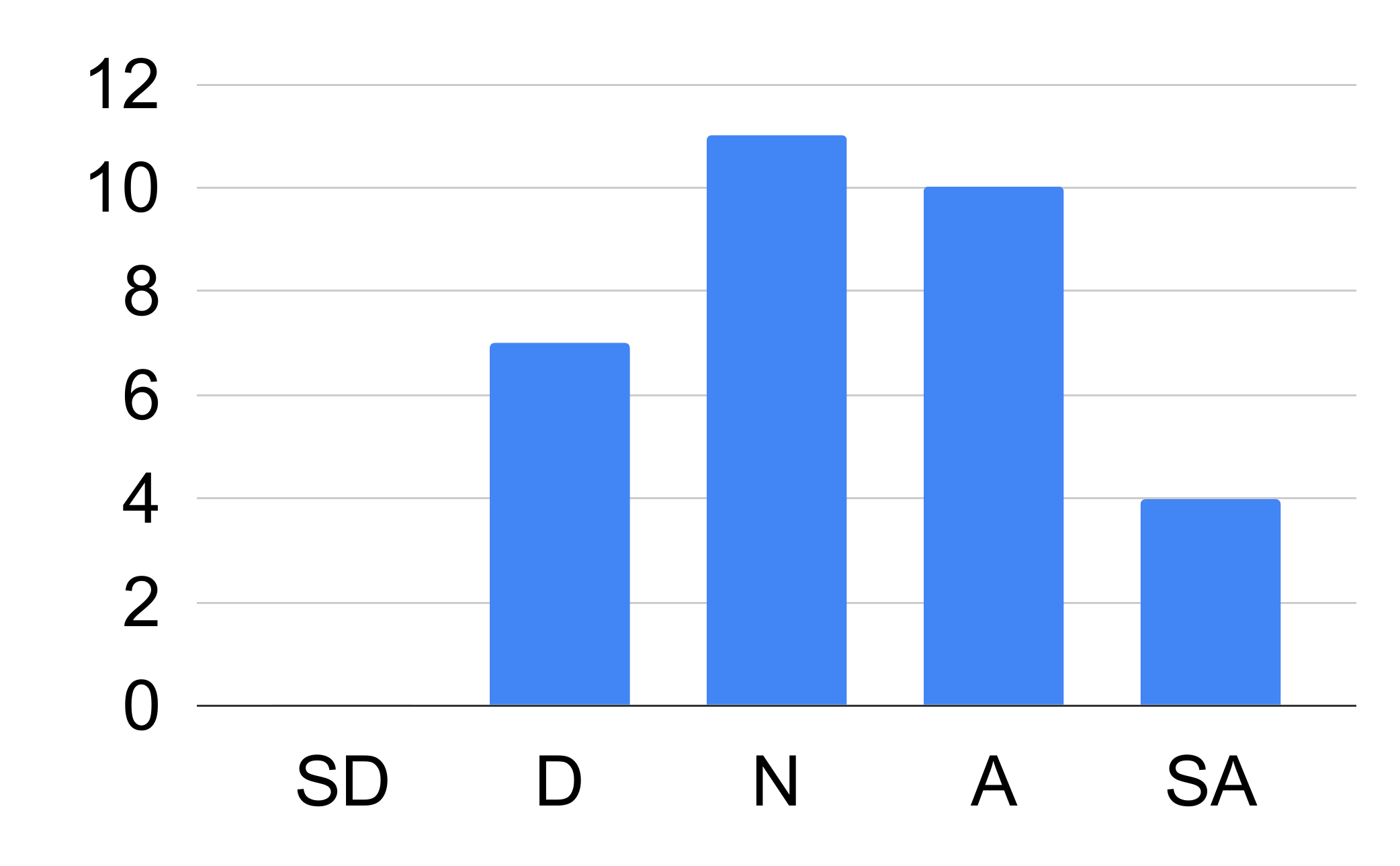}
	\end{minipage}
	\begin{minipage}{0.45\linewidth}
	    \footnotesize
		\centering
        \begin{tabular}{ | l | l | l | l | }
        \hline
        \textbf{Group} & \textbf{Median} & \textbf{Mode} & \textbf{Sim. Mode} \\ \hline
        All & N & N & A \\ \hline
        Pro./eng. & N & N & A \\ \hline
        Sol./cons. & N & A & A \\ \hline
        Tech. ed. & A & A & A \\ \hline
        Other ed. & N & N & N \\ \hline
        Exp. & A/N & N & A \\ \hline
        New & N & D & N \\ \hline
      \end{tabular}
	\end{minipage}\hfill
	\caption{Results. \emph{Process models are used as requirements specifications that are then implemented in IT systems}.}
	\label{fig:chart_12}

	\begin{minipage}{0.5\linewidth}
		\centering
		\includegraphics[trim={1.25cm 0 1.75cm 0},clip,width=55mm]{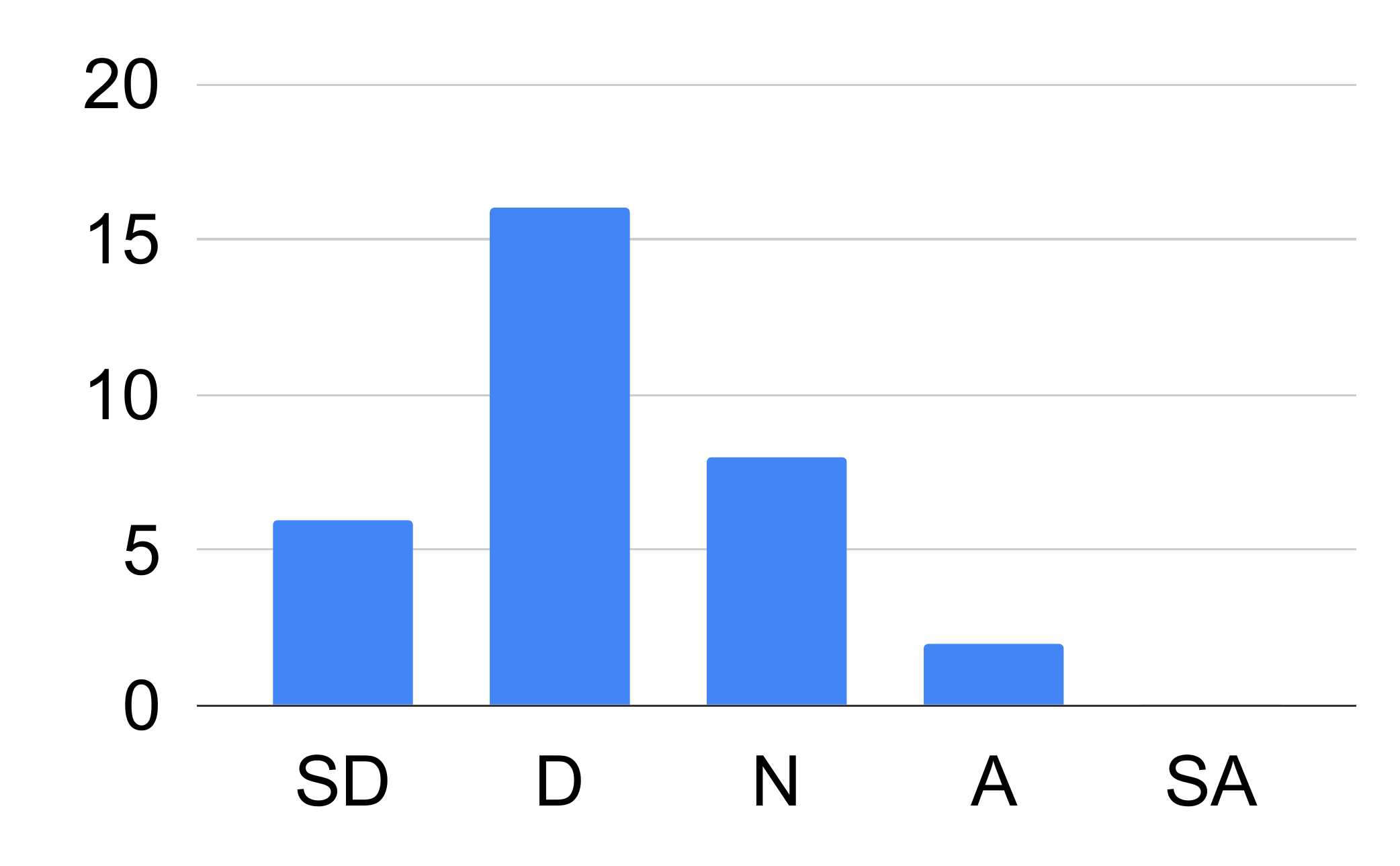}
	\end{minipage}
	\begin{minipage}{0.45\linewidth}
	    \footnotesize
		\centering
        \begin{tabular}{ | l | l | l | l | }
        \hline
        \textbf{Group} & \textbf{Median} & \textbf{Mode} & \textbf{Sim. Mode} \\ \hline
        All & D & D & D \\ \hline
        Pro./eng. & D & D & D \\ \hline
        Sol./cons. & D & D & D \\ \hline
        Tech. ed. & D & D & D \\ \hline
        Other ed. & D & D & D \\ \hline
        Exp. & D & D & D \\ \hline
        New & D & D & D \\ \hline
      \end{tabular}
	\end{minipage}\hfill
	\caption{Results. \emph{Process models are not used to implement processes.}}
	\label{fig:chart_13}

	\begin{minipage}{0.5\linewidth}
		\centering
		\includegraphics[trim={1.25cm 0 1.75cm 0},clip,width=55mm]{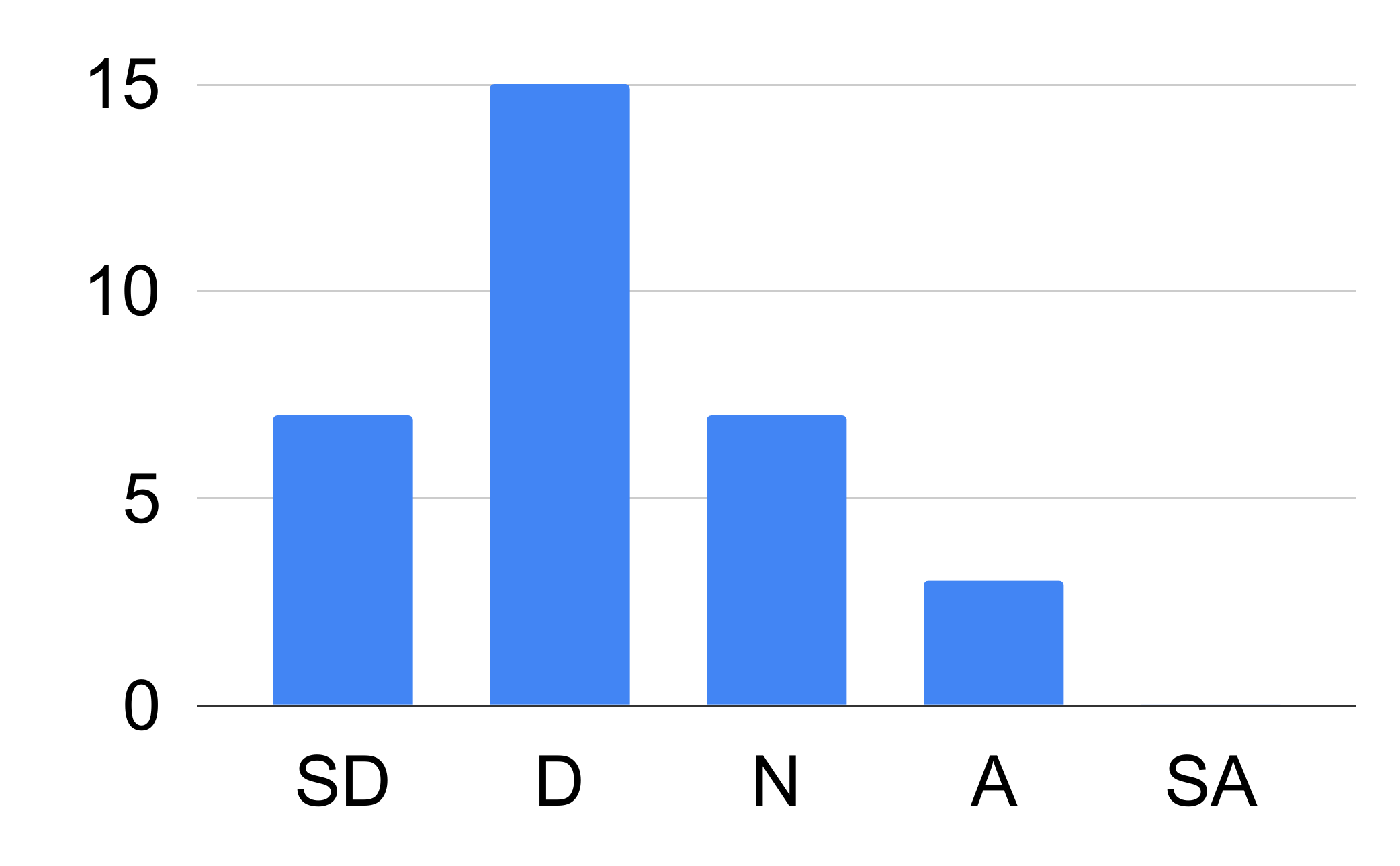}
	\end{minipage}
	\begin{minipage}{0.45\linewidth}
	    \footnotesize
		\centering
        \begin{tabular}{ | l | l | l | l | }
        \hline
        \textbf{Group} & \textbf{Median} & \textbf{Mode} & \textbf{Sim. Mode} \\ \hline
        All & D & D & D \\ \hline
        Pro./eng. & D & D & D \\ \hline
        Sol./cons. & D & D & D \\ \hline
        Tech. ed. & D & D & D \\ \hline
        Other ed. & D & D & D \\ \hline
        Exp. & D & D & D \\ \hline
        New & D & D & D \\ \hline
      \end{tabular}
	\end{minipage}\hfill
	\caption{Results. \emph{It is straightforward to find the correct scope of a process, from start to end.}}
	\label{fig:chart_21}
\end{table}
\begin{table}
	\begin{minipage}{0.5\linewidth}
		\centering
		\includegraphics[trim={1.25cm 0 1.75cm 0},clip,width=55mm]{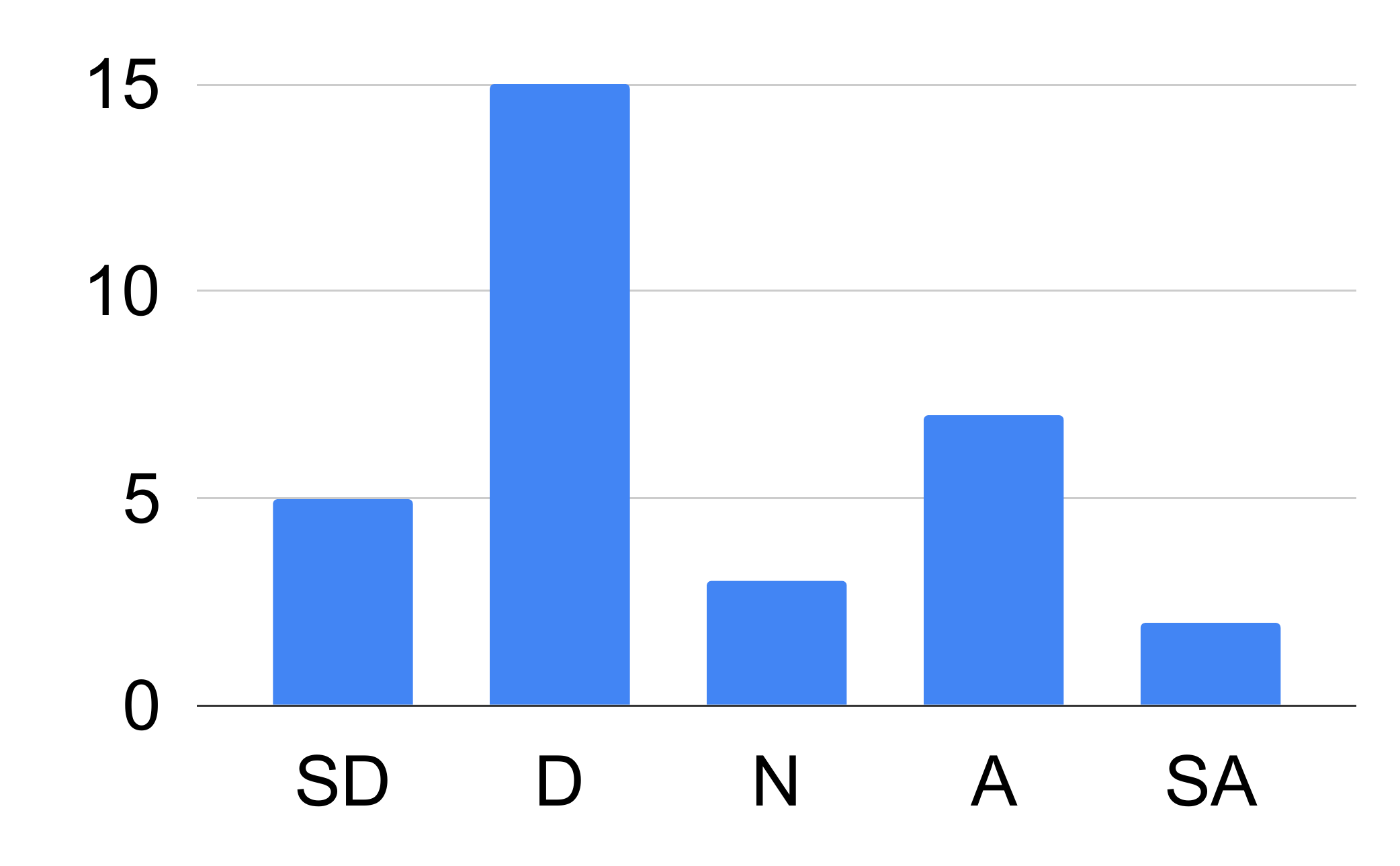}
	\end{minipage}
	\begin{minipage}{0.45\linewidth}
	    \footnotesize
		\centering
        \begin{tabular}{ | l | l | l | l | }
        \hline
        \textbf{Group} & \textbf{Median} & \textbf{Mode} & \textbf{Sim. Mode} \\ \hline
        All & D & D & D \\ \hline
        Pro./eng. & D & D & D \\ \hline
        Sol./cons. & D & D & D \\ \hline
        Tech. ed. & D & D & D \\ \hline
        Other ed. & D & D & D \\ \hline
        Exp. & D & D & D \\ \hline
        New & D & D & D \\ \hline
      \end{tabular}
	\end{minipage}\hfill
	\caption{Results. \emph{It is straightforward to group events to process instances (finding the case ID, group by case ID).}}
	\label{fig:chart_22}

	\begin{minipage}{0.5\linewidth}
		\centering
		\includegraphics[trim={1.25cm 0 1.75cm 0},clip,width=55mm]{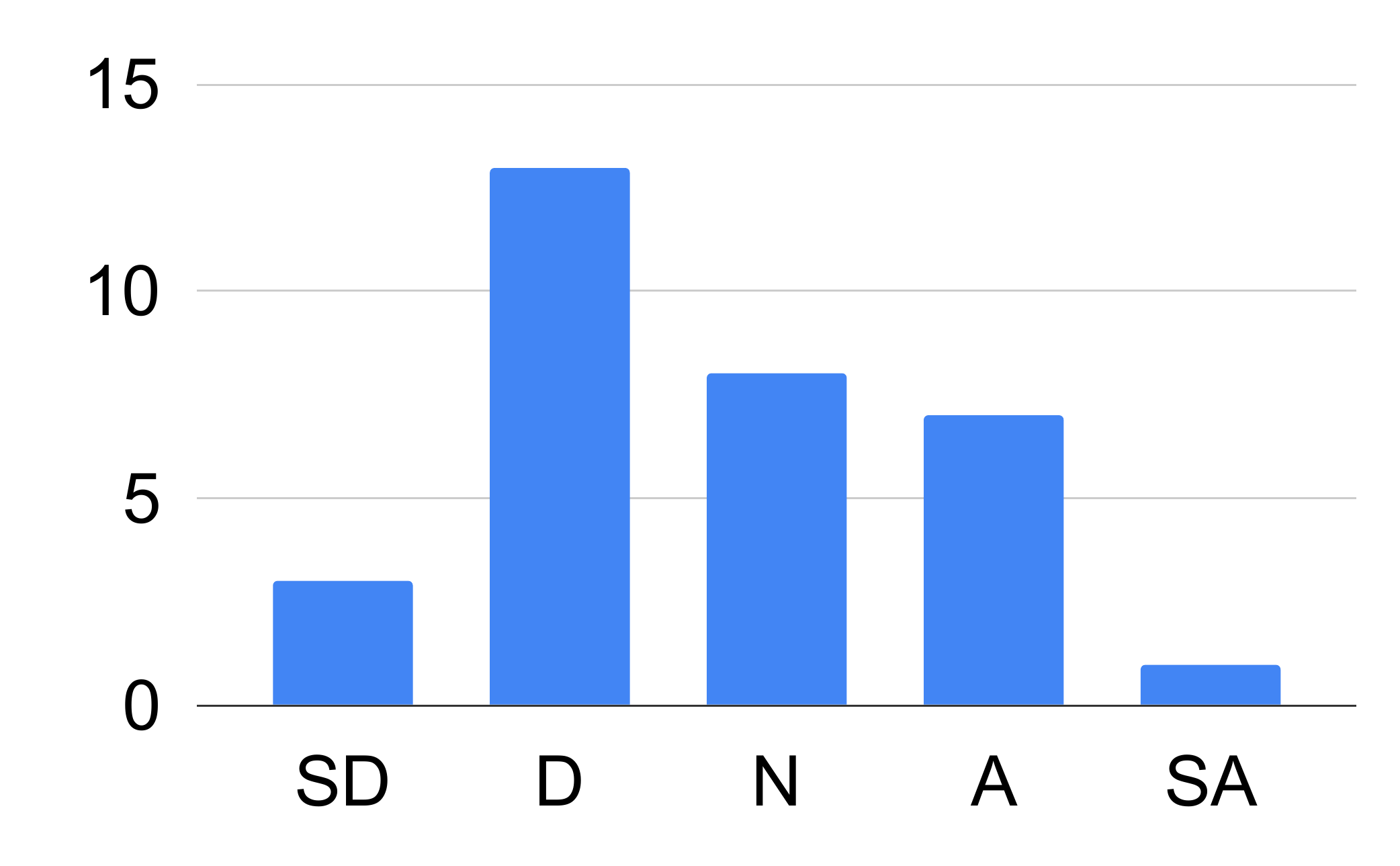}
	\end{minipage}
	\begin{minipage}{0.45\linewidth}
	    \footnotesize
		\centering
        \begin{tabular}{ | l | l | l | l | }
        \hline
        \textbf{Group} & \textbf{Median} & \textbf{Mode} & \textbf{Sim. Mode} \\ \hline
        All & D/N & D & D \\ \hline
        Pro./eng. & D & D & D \\ \hline
        Sol./cons. & N & D & D \\ \hline
        Tech. ed. & N & D & D \\ \hline
        Other ed. & D & D & D \\ \hline
        Exp. & N & D & D \\ \hline
        New & D & D & D \\ \hline
      \end{tabular}
	\end{minipage}\hfill
	\caption{Results. \emph{It is straightforward to locate the data sources that we need for generating an event log.}}
	\label{fig:chart_23}

	\begin{minipage}{0.5\linewidth}
		\centering
		\includegraphics[trim={1.25cm 0 1.75cm 0},clip,width=55mm]{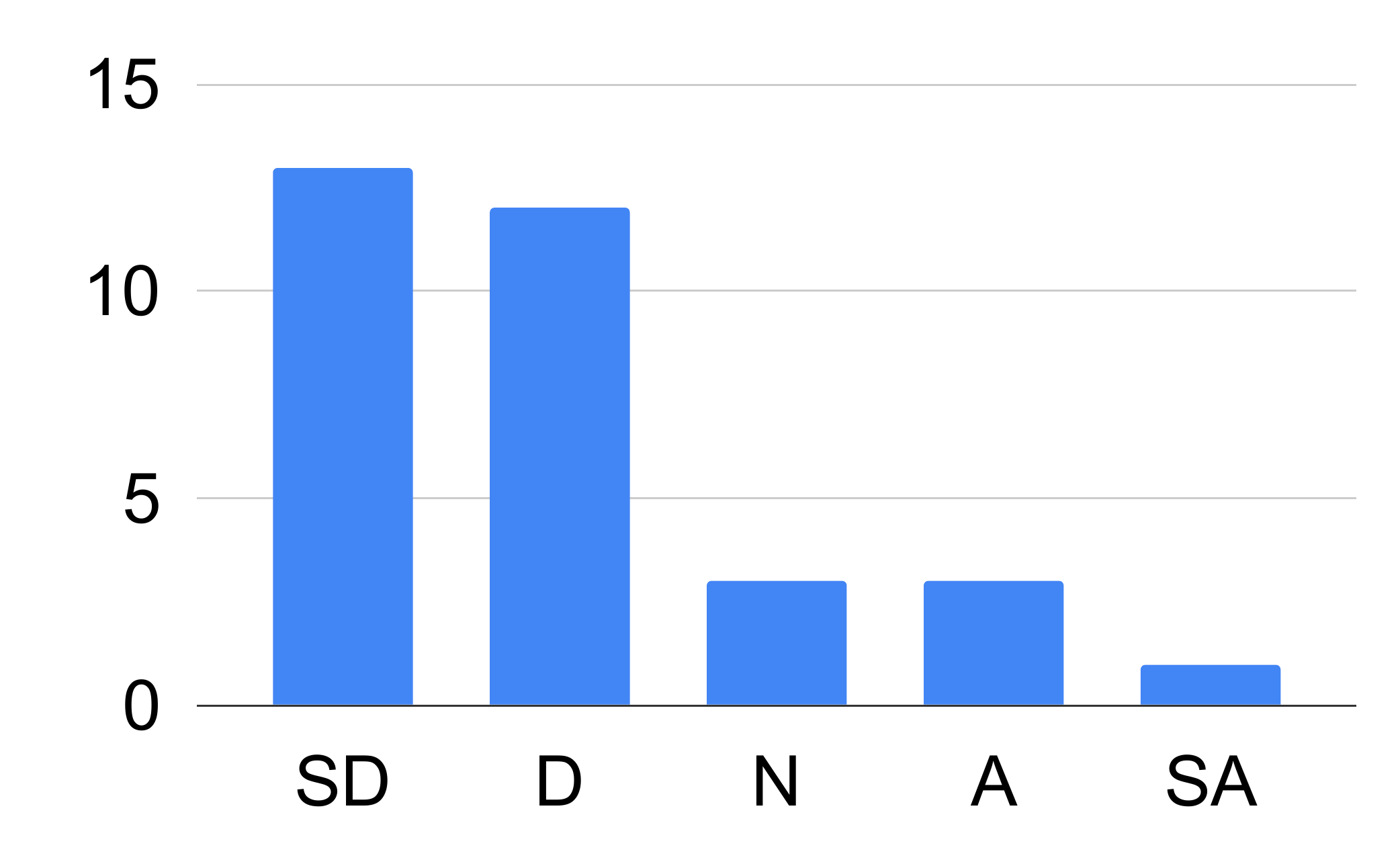}
	\end{minipage}
	\begin{minipage}{0.45\linewidth}
	    \footnotesize
		\centering
        \begin{tabular}{ | l | l | l | l | }
        \hline
        \textbf{Group} & \textbf{Median} & \textbf{Mode} & \textbf{Sim. Mode} \\ \hline
        All & D & SD & D \\ \hline
        Pro./eng. & D & SD & D \\ \hline
        Sol./cons. & D & SD & D \\ \hline
        Tech. ed. & D & SD & D \\ \hline
        Other ed. & D & SD & D \\ \hline
        Exp. & D & SD & D \\ \hline
        New & D & D & D \\ \hline
      \end{tabular}
	\end{minipage}\hfill
	\caption{Results. \emph{Typically, data quality issues do not affect event log generation.}}
	\label{fig:chart_24}

	\begin{minipage}{0.5\linewidth}
		\centering
		\includegraphics[trim={1.25cm 0 1.75cm 0},clip,width=55mm]{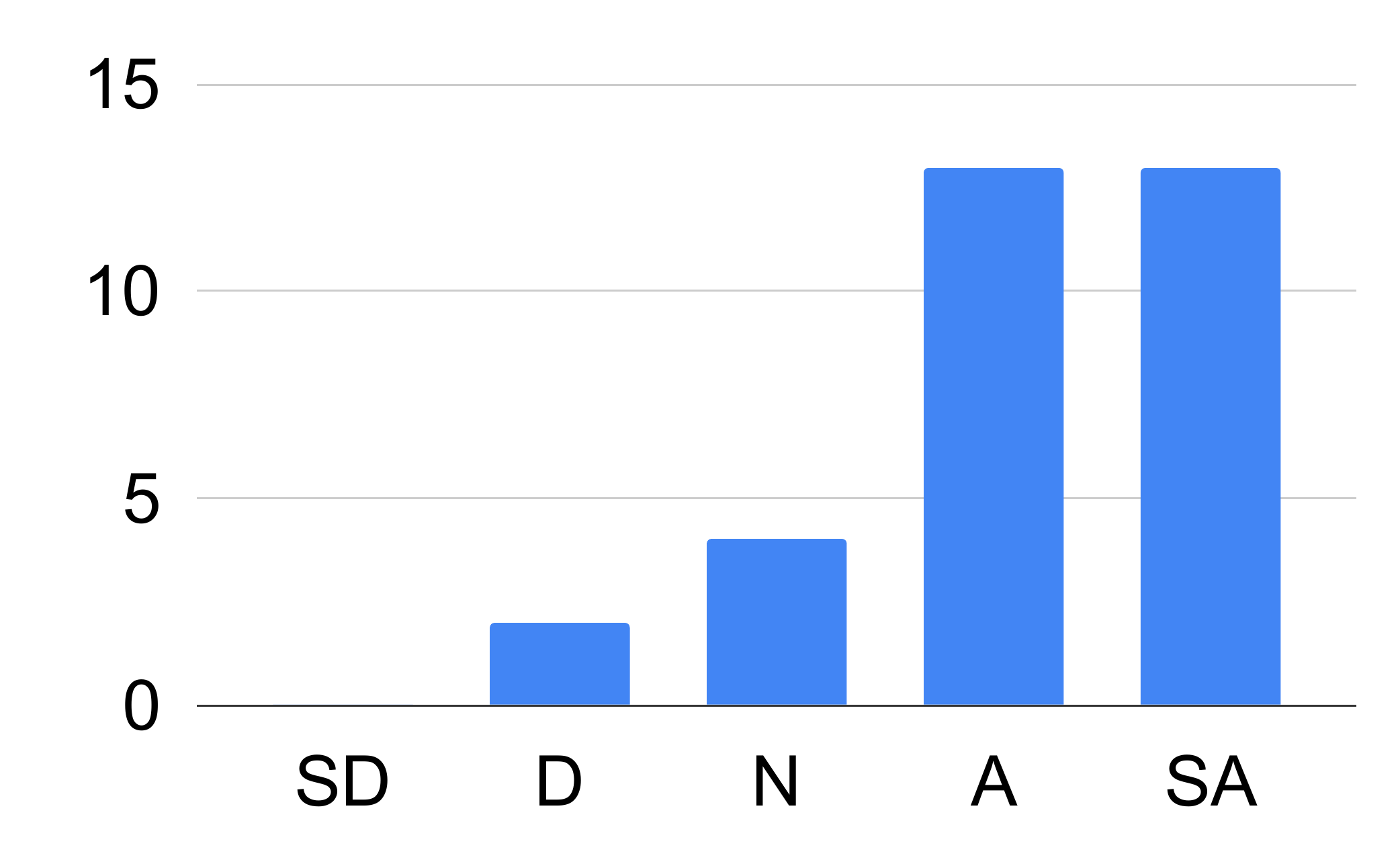}
	\end{minipage}
	\begin{minipage}{0.45\linewidth}
	    \footnotesize
		\centering
        \begin{tabular}{ | l | l | l | l | }
        \hline
        \textbf{Group} & \textbf{Median} & \textbf{Mode} & \textbf{Sim. Mode} \\ \hline
        All & A & SA/A & A \\ \hline
        Pro./eng. & A & SA & A \\ \hline
        Sol./cons. & A & A & A \\ \hline
        Tech. ed. & A & SA & A \\ \hline
        Other ed. & A & A & A \\ \hline
        Exp. & A & A & A \\ \hline
        New & A & SA & A \\ \hline
      \end{tabular}
	\end{minipage}\hfill
	\caption{Results. \emph{Event log generation incurs significant efforts in process mining projects.}}
	\label{fig:chart_31}
\end{table}
\begin{table}
	\begin{minipage}{0.5\linewidth}
		\centering
		\includegraphics[trim={1.25cm 0 1.75cm 0},clip,width=55mm]{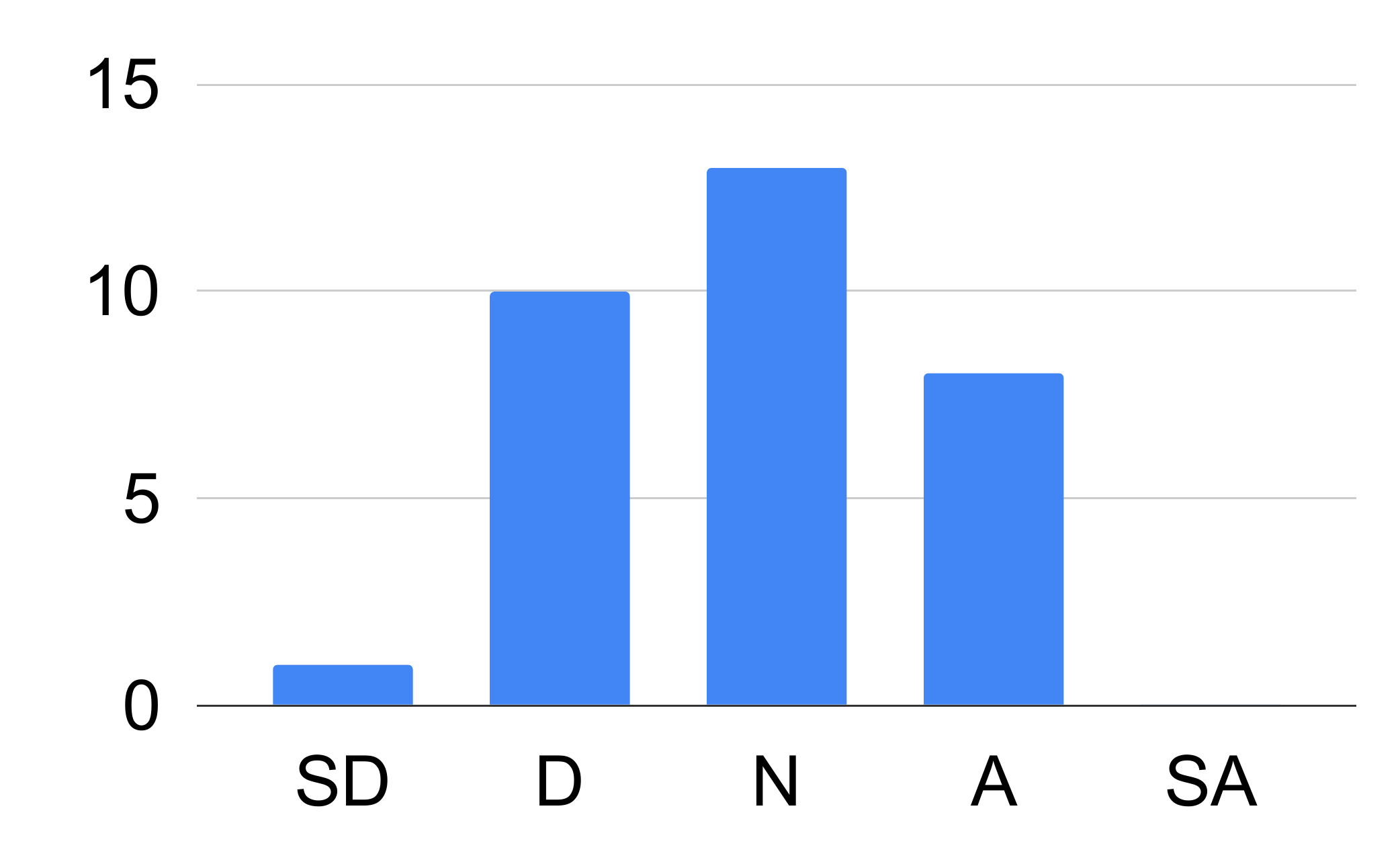}
	\end{minipage}
	\begin{minipage}{0.45\linewidth}
	    \footnotesize
		\centering
        \begin{tabular}{ | l | l | l | l | }
        \hline
        \textbf{Group} & \textbf{Median} & \textbf{Mode} & \textbf{Sim. Mode} \\ \hline
        All & N & N & N \\ \hline
        Pro./eng. & N & D & D/N \\ \hline
        Sol./cons. & N & N & N \\ \hline
        Tech. ed. & N & N & N \\ \hline
        Other ed. & N & D & D \\ \hline
        Exp. & N & N & N \\ \hline
        New & N & N & N \\ \hline
      \end{tabular}
	\end{minipage}\hfill
	\caption{Results. \emph{Extract-Transform-Load (ETL) pipelines provide all information needed in an event log.}}
	\label{fig:chart_32}
	\begin{minipage}{0.5\linewidth}
		\centering
		\includegraphics[trim={1.25cm 0 1.75cm 0},clip,width=55mm]{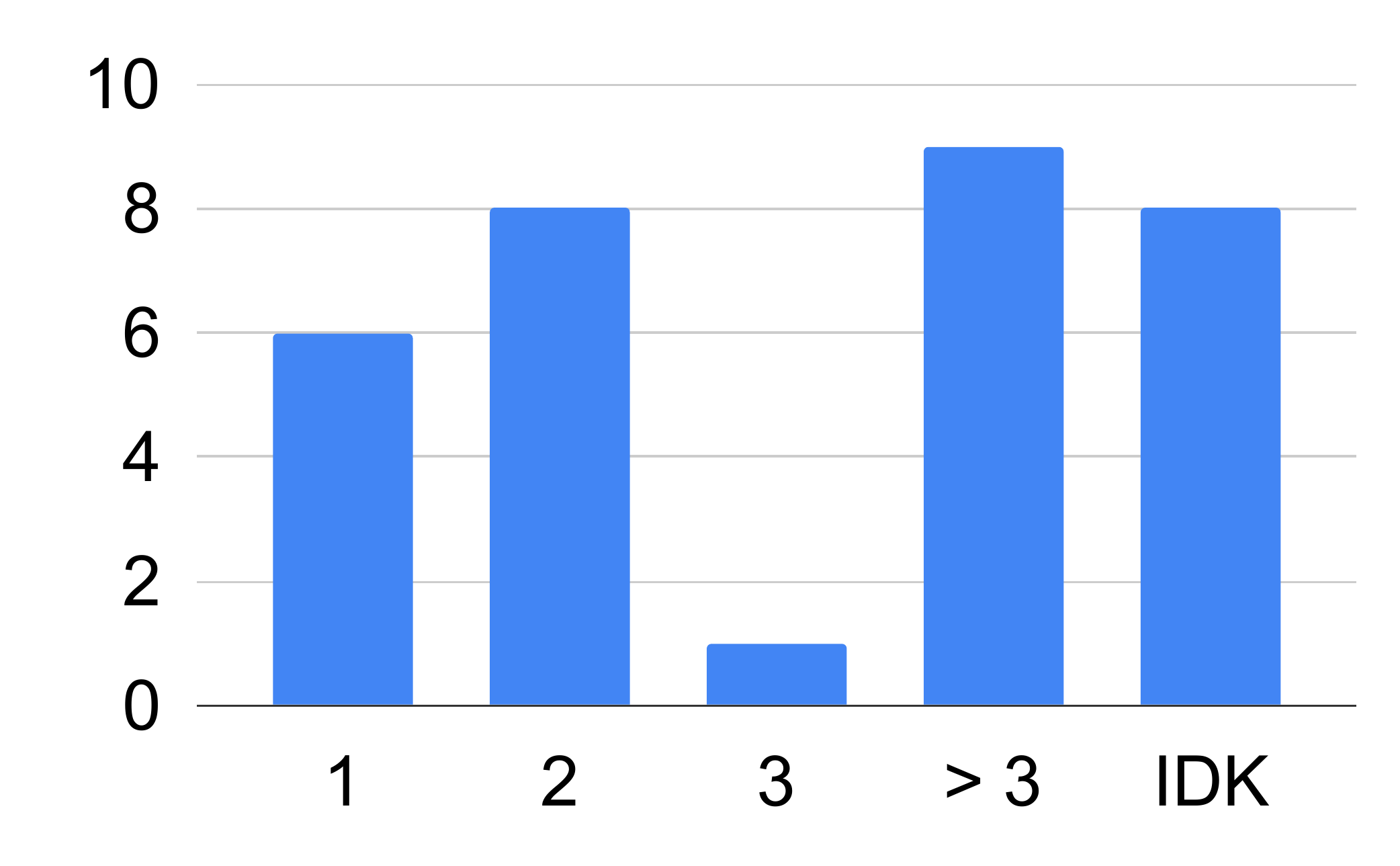}
	\end{minipage}
	\begin{minipage}{0.45\linewidth}
	    \footnotesize
		\centering
        \begin{tabular}{ | l | l | }
        \hline
        \textbf{Group} & \textbf{Mode} \\ \hline
        All & More than 3 \\ \hline
        Pro./eng. & More than 3 \\ \hline
        Sol./cons. & 1, 2  \\ \hline
        Tech. ed. & 1, ``I don't know''\\ \hline
        Other ed. & More than 3 \\ \hline
        Exp. & 2 \\ \hline
        New & More than 3 \\ \hline
      \end{tabular}
	\end{minipage}\hfill
	\caption{Results. \emph{How many backend systems are typically providing the data for a single event log?}}
	\label{fig:chart_33}
	\begin{minipage}{0.5\linewidth}
		\centering
		\includegraphics[trim={1.25cm 0 1.75cm 0},clip,width=55mm]{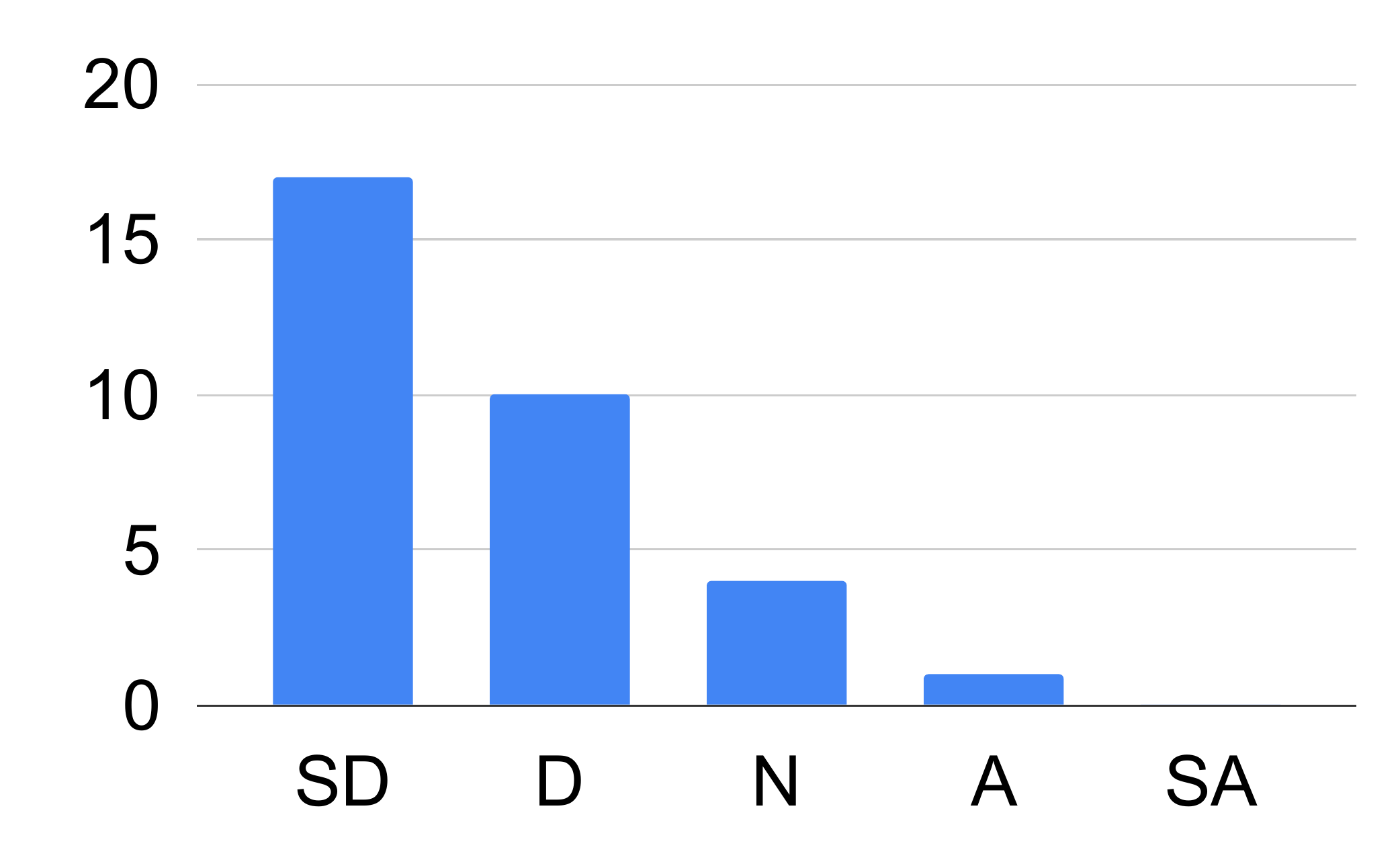}
	\end{minipage}
	\begin{minipage}{0.45\linewidth}
	    \footnotesize
		\centering
        \begin{tabular}{ | l | l | l | l | }
        \hline
        \textbf{Group} & \textbf{Median} & \textbf{Mode} & \textbf{Sim. Mode} \\ \hline
        All & SD & SD & D \\ \hline
        Pro./eng. & D & SD & D \\ \hline
        Sol./cons. & SD & SD & D \\ \hline
        Tech. ed. & SD & SD & D \\ \hline
        Other ed. & D & D & D \\ \hline
        Exp. & SD/D & SD & D \\ \hline
        New & SD & SD & D \\ \hline
      \end{tabular}
	\end{minipage}\hfill
	\caption{Results. \emph{In a given system, the information needed for an event log is stored in a single relational table.}}
	\label{fig:chart_34}
	\begin{minipage}{0.5\linewidth}
		\centering
		\includegraphics[trim={1.25cm 0 1.75cm 0},clip,width=55mm]{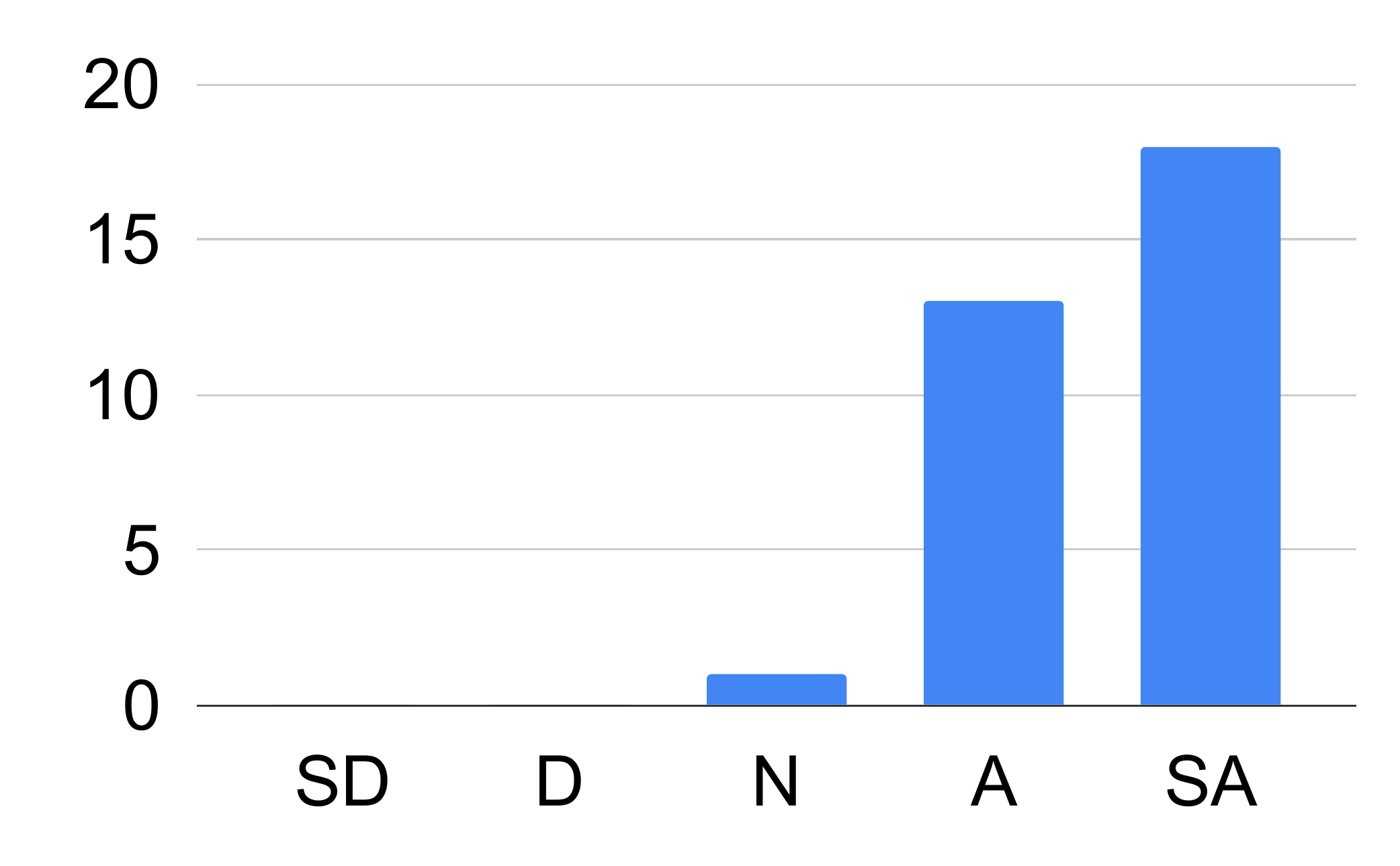}
	\end{minipage}
	\begin{minipage}{0.45\linewidth}
	    \footnotesize
		\centering
        \begin{tabular}{ | l | l | l | l | }
        \hline
        \textbf{Group} & \textbf{Median} & \textbf{Mode} & \textbf{Sim. Mode} \\ \hline
        All & SA & SA & A \\ \hline
        Pro./eng. & A & A & A \\ \hline
        Sol./cons. & SA & SA & A \\ \hline
        Tech. ed. & SA & SA & A \\ \hline
        Other ed. & SA & SA & A \\ \hline
        Exp. & SA & SA & A \\ \hline
        New & SA/A & SA/A & A \\ \hline
      \end{tabular}
	\end{minipage}\hfill
	\caption{Results. \emph{In process discovery, we encounter complex process models.}}
	\label{fig:chart_41}
\end{table}
\begin{table}
	\begin{minipage}{0.5\linewidth}
		\centering
		\includegraphics[trim={1.25cm 0 1.75cm 0},clip,width=55mm]{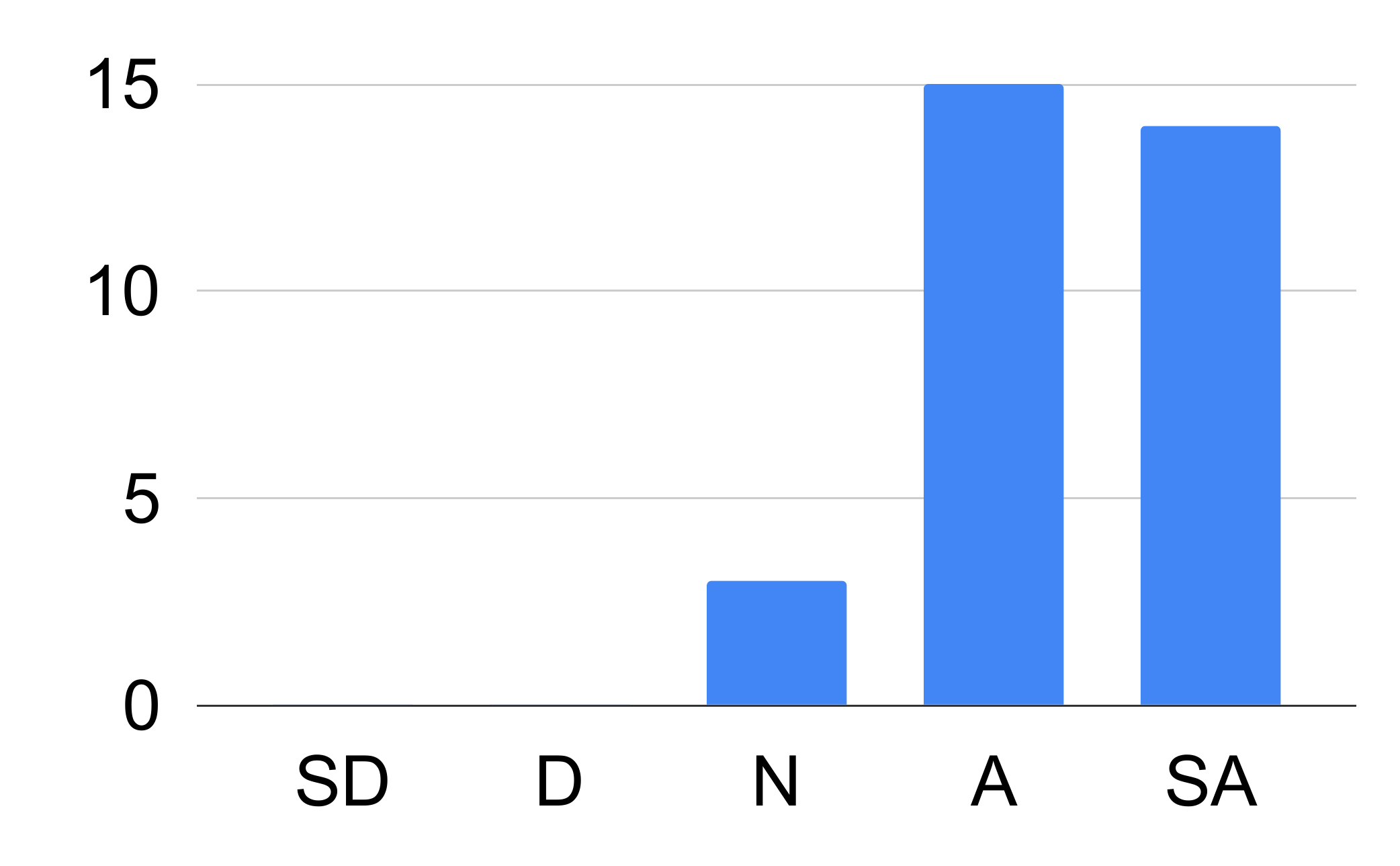}
	\end{minipage}
	\begin{minipage}{0.45\linewidth}
	    \footnotesize
		\centering
        \begin{tabular}{ | l | l | l | l | }
        \hline
        \textbf{Group} & \textbf{Median} & \textbf{Mode} & \textbf{Sim. Mode} \\ \hline
        All & A & A & A \\ \hline
        Pro./eng. & A & A & A \\ \hline
        Sol./cons. & SA & SA & A \\ \hline
        Tech. ed. & SA & SA & A \\ \hline
        Other ed. & A & A & A \\ \hline
        Exp. & SA & SA & A \\ \hline
        New & A & A & A \\ \hline
      \end{tabular}
	\end{minipage}\hfill
	\caption{Results. \emph{In process discovery, the ordering of activities is important to me (e.g., it is important to know that ``activity A always precedes activity B'').}}
	\label{fig:chart_42}
	\begin{minipage}{0.5\linewidth}
		\centering
		\includegraphics[trim={1.25cm 0 1.75cm 0},clip,width=55mm]{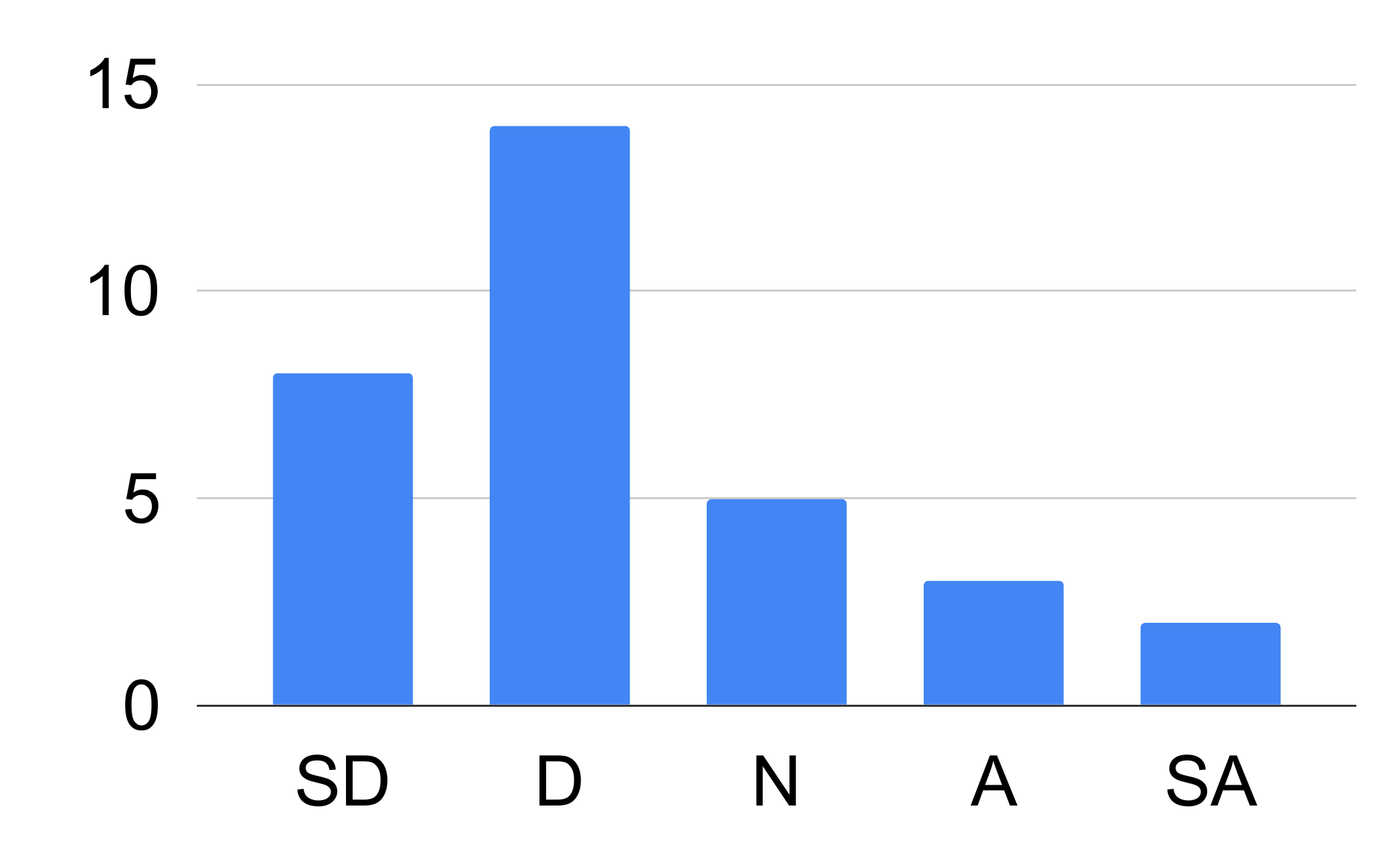}
	\end{minipage}
	\begin{minipage}{0.45\linewidth}
	    \footnotesize
		\centering
        \begin{tabular}{ | l | l | l | l | }
        \hline
        \textbf{Group} & \textbf{Median} & \textbf{Mode} & \textbf{Sim. Mode} \\ \hline
        All & D & D & D \\ \hline
        Pro./eng. & D & D & D \\ \hline
        Sol./cons. & N & D & D \\ \hline
        Tech. ed. & D & D & D \\ \hline
        Other ed. & D & SD & D \\ \hline
        Exp. & D & SD & D \\ \hline
        New & D & D & D \\ \hline
      \end{tabular}
	\end{minipage}\hfill
	\caption{Results. \emph{Most discovered processes are sequential (no branching or concurrency).}}
	\label{fig:chart_43}

	\begin{minipage}{0.5\linewidth}
		\centering
		\includegraphics[trim={1.25cm 0 1.75cm 0},clip,width=55mm]{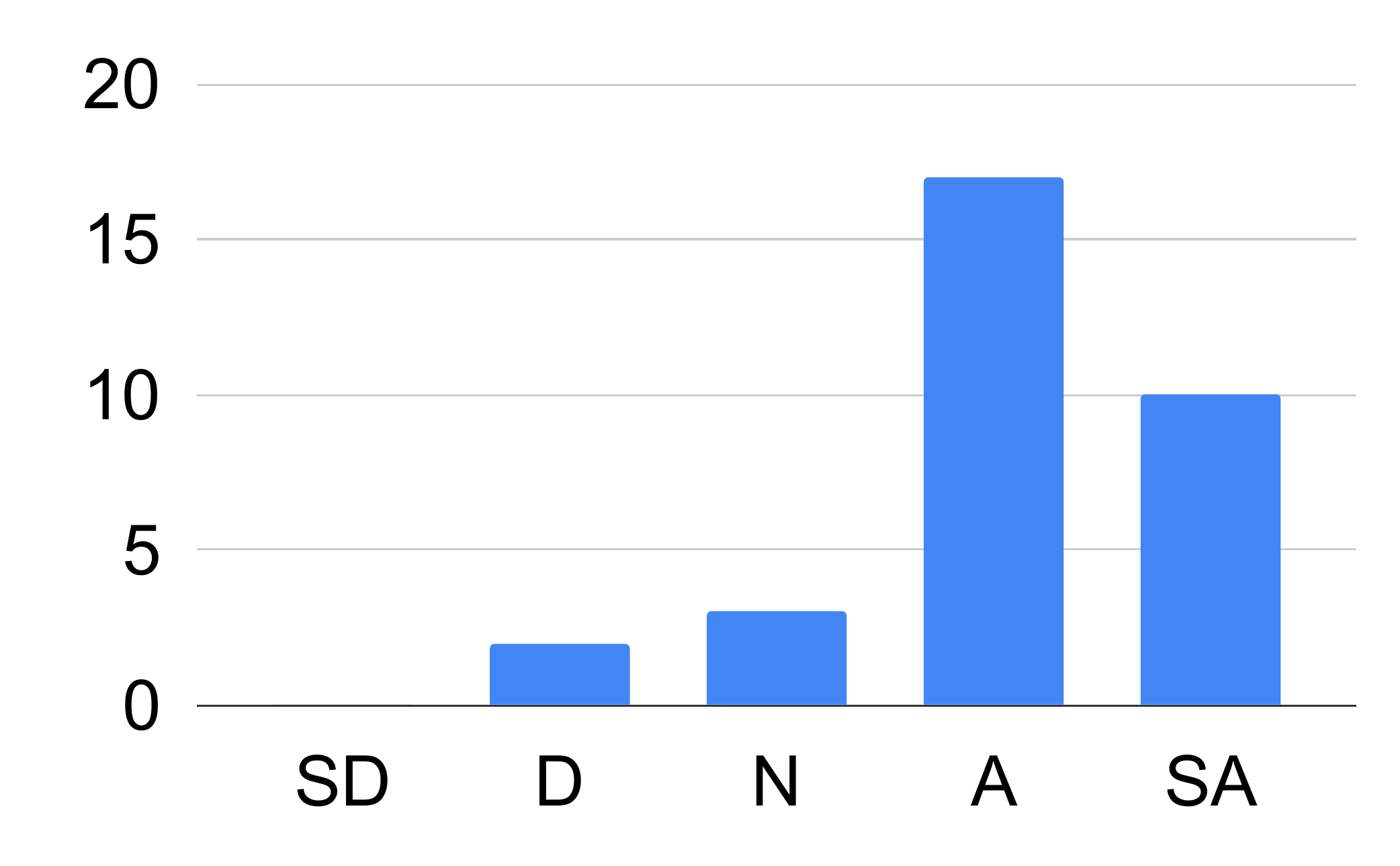}
	\end{minipage}
	\begin{minipage}{0.45\linewidth}
	    \footnotesize
		\centering
        \begin{tabular}{ | l | l | l | l | }
        \hline
        \textbf{Group} & \textbf{Median} & \textbf{Mode} & \textbf{Sim. Mode} \\ \hline
        All & A & A & A \\ \hline
        Pro./eng. & A & A & A \\ \hline
        Sol./cons. & A & A & A \\ \hline
        Tech. ed. & A & A & A \\ \hline
        Other ed. & A & A & A \\ \hline
        Exp. & A & A & A \\ \hline
        New & A & A & A \\ \hline
      \end{tabular}
	\end{minipage}\hfill
	\caption{Results. \emph{An important goal of process mining is the calculation of KPIs.}}
	\label{fig:chart_51}
	\begin{minipage}{0.5\linewidth}
		\centering
		\includegraphics[trim={1.25cm 0 1.75cm 0},clip,width=55mm]{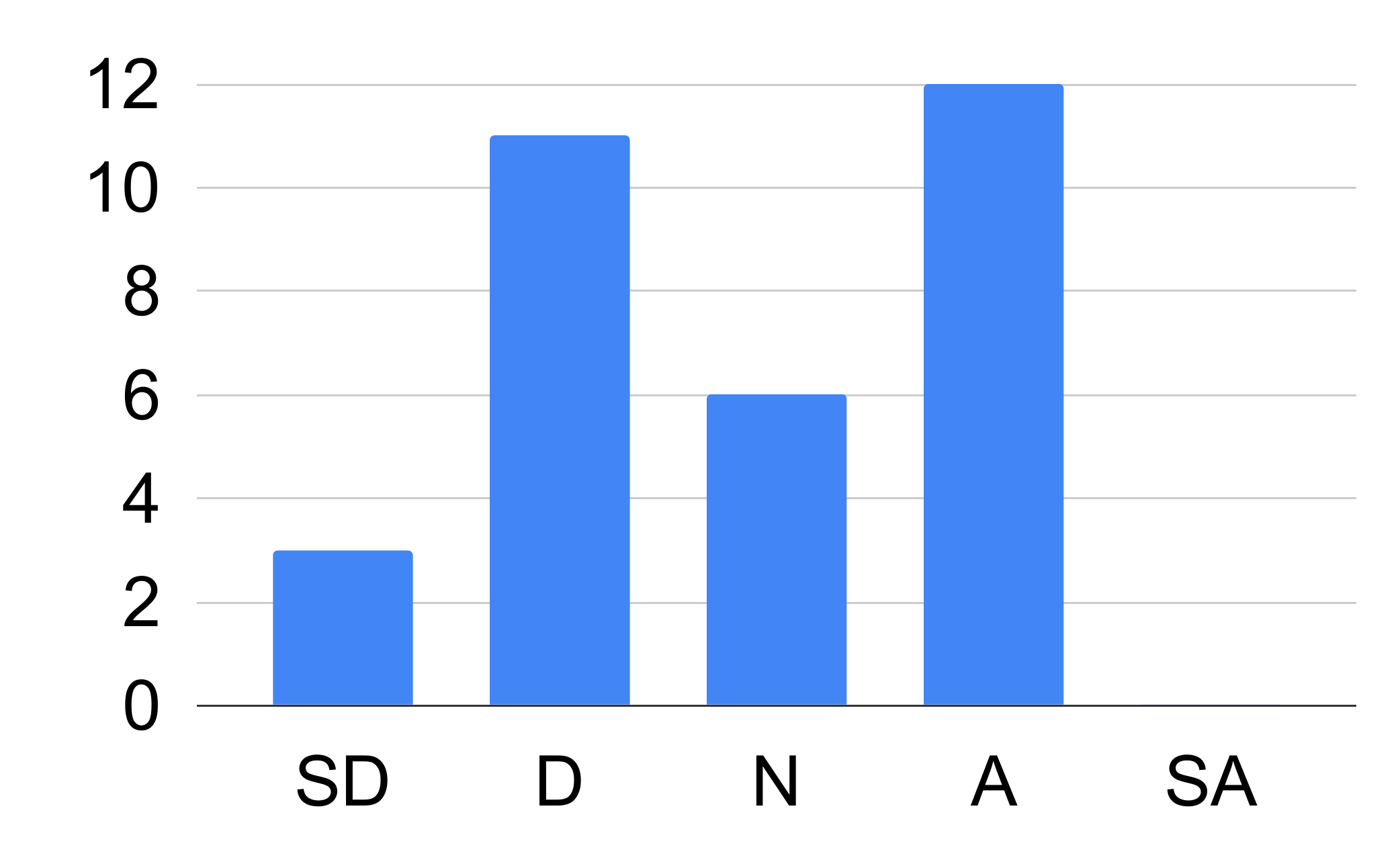}
	\end{minipage}
	\begin{minipage}{0.45\linewidth}
	    \footnotesize
		\centering
        \begin{tabular}{ | l | l | l | l | }
        \hline
        \textbf{Group} & \textbf{Median} & \textbf{Mode} & \textbf{Sim. Mode} \\ \hline
        All & N & A & D \\ \hline
        Pro./eng. & D & D & D \\ \hline
        Sol./cons. & N & A & A \\ \hline
        Tech. ed. & N & A & A \\ \hline
        Other ed. & D & D & D \\ \hline
        Exp. & D & D & D \\ \hline
        New & N & A & A \\ \hline
      \end{tabular}
	\end{minipage}\hfill
	\caption{Results. \emph{In process mining, it is difficult to identify meaningful KPIs.}}
	\label{fig:chart_52}
\end{table}
\begin{table}
	\begin{minipage}{0.5\linewidth}
		\centering
		\includegraphics[trim={1.25cm 0 1.75cm 0},clip,width=55mm]{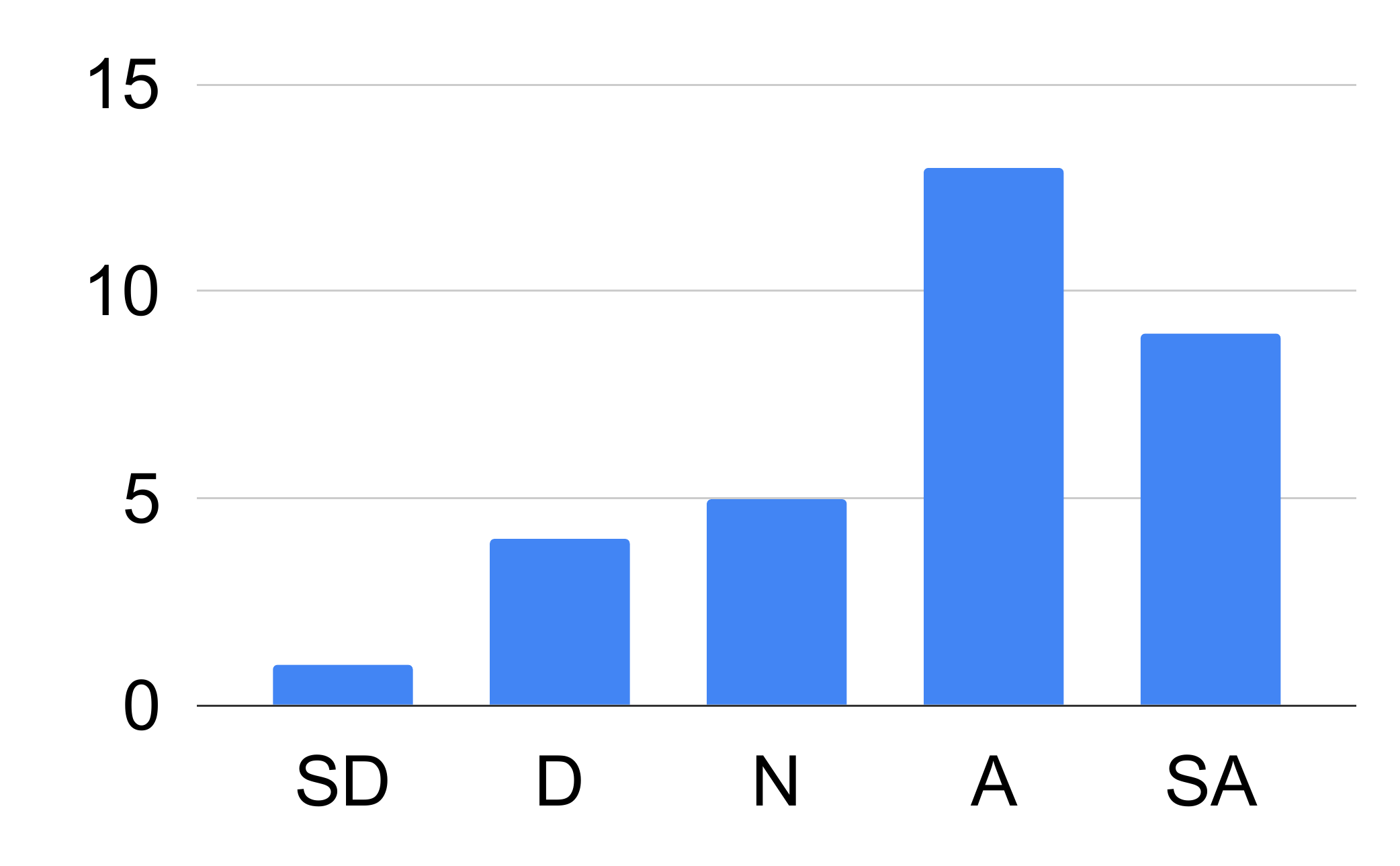}
	\end{minipage}
	\begin{minipage}{0.45\linewidth}
	    \footnotesize
		\centering
        \begin{tabular}{ | l | l | l | l | }
        \hline
        \textbf{Group} & \textbf{Median} & \textbf{Mode} & \textbf{Sim. Mode} \\ \hline
        All & A & A & A \\ \hline
        Pro./eng. & A & A & A \\ \hline
        Sol./cons. & A & A & A \\ \hline
        Tech. ed. & A & A & A \\ \hline
        Other ed. & A & SA & A \\ \hline
        Exp. & A & A & A \\ \hline
        New & A & A & A \\ \hline
      \end{tabular}
	\end{minipage}\hfill
	\caption{Results. \emph{Comparing the event log with a process model is important in process analysis.}}
	\label{fig:chart_53}
	\begin{minipage}{0.5\linewidth}
		\centering
		\includegraphics[trim={1.25cm 0 1.75cm 0},clip,width=55mm]{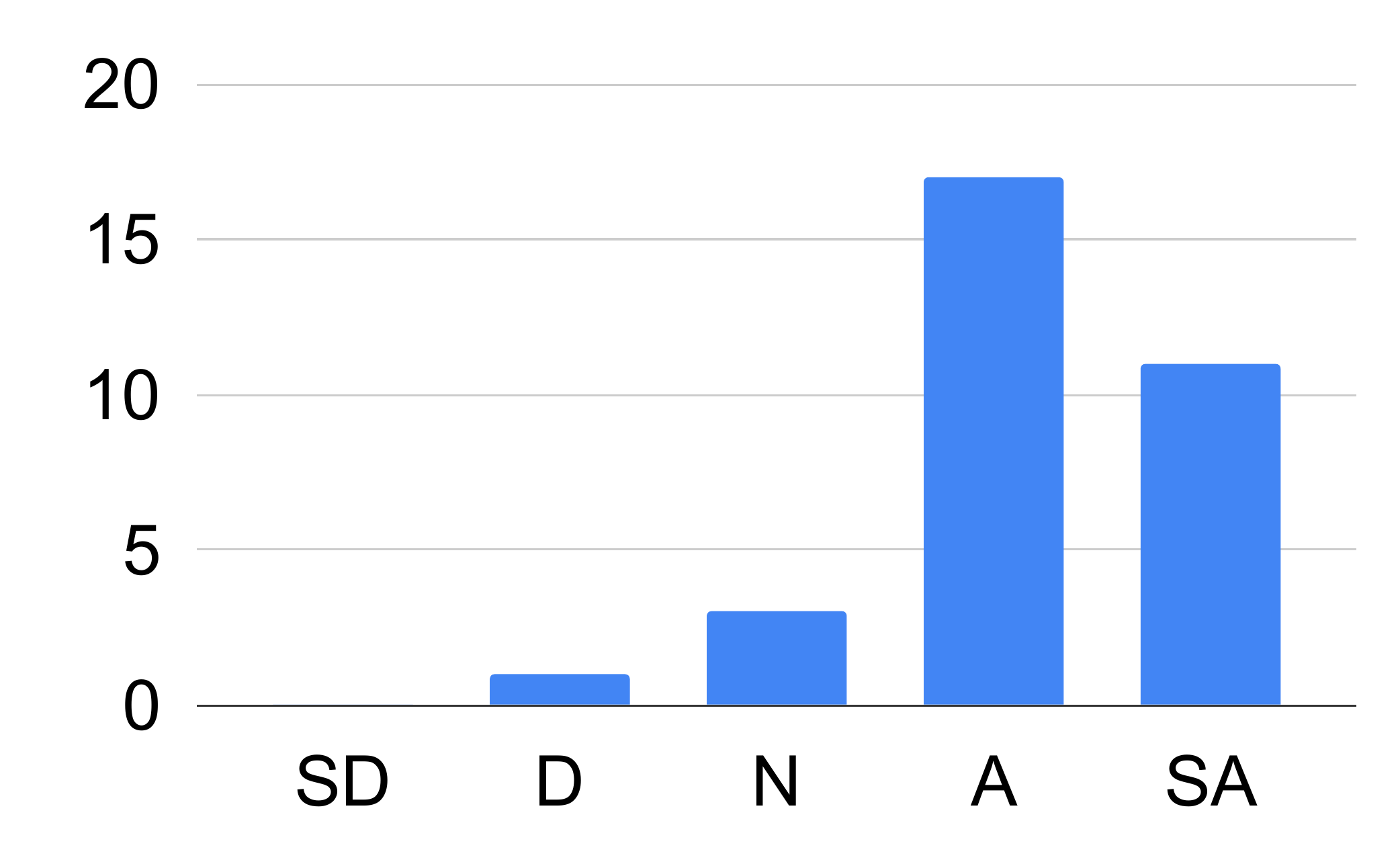}
	\end{minipage}
	\begin{minipage}{0.45\linewidth}
	    \footnotesize
		\centering
        \begin{tabular}{ | l | l | l | l | }
        \hline
        \textbf{Group} & \textbf{Median} & \textbf{Mode} & \textbf{Sim. Mode} \\ \hline
        All & A & A & A \\ \hline
        Pro./eng. & A & SA/A & A \\ \hline
        Sol./cons. & A & A & A \\ \hline
        Tech. ed. & A & A & A \\ \hline
        Other ed. & A & SA/A & A \\ \hline
        Exp. & SA/A & SA & A \\ \hline
        New & A & A & A \\ \hline
      \end{tabular}
	\end{minipage}\hfill
	\caption{Results. \emph{A better integration of Business Intelligence (BI) and process mining would be valuable.}}
	\label{fig:chart_54}
\end{table}
%
%
%
\end{document}